\pdfoutput=1
\documentclass[11pt,twoside,a4paper,cmspaper,final,collab]{cms-tdr}
\def\svnVersion{1b6515d}\def\svnDate{2026/10/06}\def\cmsCernNoTag{CERN-EP-2026-116}\def\cmsCernDate{\today}\def\cmsMessage{Published in the Journal of High Energy Physics as \href{http://dx.doi.org/10.1007/JHEP10(2026)006}{\doi{10.1007/JHEP10(2026)006}.}}
\begin{document}\cmsNoteHeader{HIG-24-002}

\cmsNoteHeader{HIG-24-002}

\newcommand{\fvbf}{f_\text{VBF}}
\newcommand{\mreco}{m_{4\Pell}^{\text{reco}}}
\newcommand{\mgen}{m_{4\Pell}^{\text{gen}}}
\newcommand{\dbkg}{D_\text{bkg}^{\text{kin}}}
\newcommand{\dtwoj}{D_\text{2jet}^{\text{VBF}}}
\newcommand{\MX}{\ensuremath{m_{\PX}}}
\newcommand{\gammaX}{\ensuremath{\Gamma_{\PX}}}

\title{Search for a new heavy scalar resonance decaying to a pair of \texorpdfstring{\PZ}{Z} bosons in the four-lepton final state in proton-proton collisions at \texorpdfstring{$\sqrt{s} = 13\TeV$}{sqrt(s) = 13 TeV}}

\date{\today}

\abstract{
   A search for a new heavy scalar resonance decaying to two \PZ bosons, each subsequently decaying to a pair of electrons or muons, is presented. The results are based on a proton-proton collision data set collected by the CMS experiment at the LHC at a center-of-mass energy of 13\TeV, corresponding to an integrated luminosity of 138\fbinv. The search is performed over a wide range of resonance masses from 130\GeV to 3\TeV, considering both narrow- and broad-width scenarios, and considering the gluon fusion and vector boson fusion production processes. For the broad-width scenario, the interference between the new resonance, the 125\GeV Higgs boson production, and the continuum background is taken into account. No significant excess with respect to the standard model background expectation is observed in the examined phase space. Upper limits at the 95\% confidence level are set on the product of the heavy scalar resonance production cross section and the branching fraction for its decay into two \PZ bosons. The exclusion limits range from 0.05--0.1\unit{pb} in the low-mass region to 0.005\unit{pb} in the high-mass region.
}

\hypersetup{
pdfauthor={CMS Collaboration},
pdftitle={Search for a new heavy scalar resonance decaying to a pair of Z bosons in the four-lepton final state in proton-proton collisions at sqrt(s) = 13 TeV},
pdfsubject={CMS},
pdfkeywords={CMS, Higgs, high mass, new scalar}}

\maketitle

\section{Introduction}

In 2012, the ATLAS and CMS Collaborations discovered a boson~\cite{ATLAS_2012, CMS_2012,CMS_2012_long} with a mass close to 125\GeV and with properties consistent with those expected for the Higgs boson (\PH) predicted by the standard model of particle physics (SM).
This discovery represents a major breakthrough for the field of particle physics and a significant validation of the SM, which postulates the existence of this particle as an excitation of the Higgs field.

The SM, however, presents several limitations. For example, it does not include a description of gravity or dark matter and does not explain the origin of the matter-antimatter asymmetry in the universe.
These issues could be addressed by a range of theories beyond the SM (BSM). In many models describing such BSM effects, additional scalar resonances are predicted, \eg, additional Higgs bosons in two-Higgs doublet models~\cite{Branco:2011iw} and radions in models with warped extra dimensions~\cite{PhysRevLett.83.3370,PhysRevLett.83.4922,carvalho2018gravity}.

Previous searches for a new boson decaying to a pair of \PZ bosons have been performed by the CMS~\cite{CMS:2015hra, CMS_high_mass_2018} and ATLAS~\cite{ATLAS:2015pre, ATLAS_high_mass_2022} Collaborations
using proton-proton collision data collected in Run~1 (2011--2012) and Run~2 (2015--2018) of the CERN LHC. In these searches, no evidence for a signal was observed.

In this paper, a search for a heavy scalar resonance \PX is presented in the decay channel $\PX\to\PZ\PZ\to4\Pell$ where the leptons \Pell are electrons (\Pe) or muons (\PGm).
Since the mass of the new resonance is not known, the search is performed over a wide mass range from 130\GeV to 3\TeV. The resonance can be produced via gluon fusion (ggF) or vector boson fusion (VBF). 
Compared with the previous CMS search at 13\TeV~\cite{CMS_high_mass_2018}, which used the 2016 data set corresponding to an integrated luminosity of 35.9\fbinv, this analysis uses the full Run~2 data set of 138\fbinv.
It also benefits from the final Run~2 reprocessing, which incorporates improved detector alignment, calibrations, and reconstruction conditions, as well as updated lepton identification and jet flavor tagging algorithms.
The numerical results are provided in HEPData~\cite{hepdata}.

\section{The CMS detector}
\label{sec:detector}

The CMS apparatus~\cite{CMS:2008xjf,CMS:2023gfb} is a multipurpose, nearly hermetic detector, designed to trigger on~\cite{CMS:2020cmk,CMS:2016ngn,CMS:2024aqx} and identify electrons, muons, photons, and charged and neutral hadrons~\cite{CMS:2020uim,CMS:2018rym,CMS:2014pgm}. Its central feature is a superconducting solenoid of 6\unit{m} internal diameter, providing a magnetic field of 3.8\unit{T}. Within the solenoid volume are a silicon pixel and strip tracker, a lead tungstate crystal electromagnetic calorimeter (ECAL), and a brass and scintillator hadron calorimeter (HCAL), each composed of a barrel and two endcap sections. Forward calorimeters extend the pseudorapidity ($\eta$) coverage provided by the barrel and endcap detectors. Muons are reconstructed using gas-ionization detectors interleaved with the layers of the steel flux-return yoke outside the solenoid. More detailed descriptions of the CMS detector, together with a definition of the coordinate system used and the relevant kinematic variables, can be found in Refs.~\cite{CMS:2008xjf,CMS:2023gfb}.

Events of interest are selected using a two-tiered trigger system. The first level (L1), composed of custom hardware processors, uses information from the calorimeters and muon detectors to select events at a rate of around 100\unit{kHz} within a fixed latency of 4\mus~\cite{CMS:2020cmk}. The second level, known as the high-level trigger, consists of a farm of processors running a version of the full event reconstruction software optimized for fast processing and reduces the event rate to a few kHz before data storage~\cite{CMS:2016ngn,CMS:2024aqx}.

\section{Data and simulated samples}
\label{sec:dataset}
The search is performed using the $\Pp\Pp$ collision data set collected by the CMS experiment in 2016--2018~\cite{CMS:2016lumi, CMS:2017lumi, CMS:2018lumi}, with a total integrated luminosity of 138\fbinv.

Events are selected by the online trigger system before undergoing the reconstruction described in Section~\ref{sec:selection}. The online selection is based on dielectron, dimuon, and electron-muon trigger algorithms. Additional triggers, requiring three leptons with low transverse momentum (\pt) thresholds and no isolation criteria, as well as single-lepton triggers, are also used to increase the selection efficiency.
The applied \pt thresholds range from 20 to 25\GeV for single-lepton triggers, from 8 to 23\GeV for double-lepton triggers, and from 5 to 16\GeV for triple-lepton triggers.

The trigger efficiency measurement follows the strategy adopted in Ref.~\cite{CMS_HIG_19_001}, using a ``tag-and-probe'' method~\cite{TnP}. The efficiency is greater than 99\% in most of the phase space and is compatible between the observed data and simulation within statistical uncertainties.

Monte Carlo (MC) simulated samples are used to model the signal, background, and interference processes.

Samples of \PX resonances produced via ggF or VBF are generated at next-to-leading order (NLO) in perturbative quantum chromodynamics (pQCD) with the \POWHEG 2.0~\cite{Alioli:2008gx,Nason:2004rx,Frixione:2007vw} generator. Their decays into two \PZ bosons and then into four leptons are modeled with \textsc{JHUGen} 7.0.2~\cite{Gao:2010qx,Bolognesi:2012mm,Anderson:2013afp,Gritsan:2016hjl}, including corrections for the $\PZ\PZ$ branching fraction and angular correlations among the fermions. Samples are generated for a wide range of masses of the \PX resonance ($\MX$), from 125\GeV to 3\TeV, with the width ($\gammaX$) set to the values predicted by the complex pole scheme \cite{GORIA2012530,PASSARINO201077} for $\MX<1\TeV$. For those with $\MX>1\TeV$, the width cannot be predicted by this scheme, and is therefore set conventionally to $0.5\MX$. 
However, the search is performed using a range of width assumptions, not restricted to the widths in the simulated samples. These samples are used to derive models for signal processes, as well as for 125\GeV Higgs boson production. 

Events in ggF production samples generated with \POWHEG are weighted to take advantage of predictions at next-to-NLO (NNLO) accuracy in pQCD, matched to a parton shower~\cite{Hamilton2013} (NNLOPS). These weights are obtained from the ratio of the production cross section at NNLOPS accuracy to the production cross section obtained with the \POWHEG event generator. The weights are applied as a function of the number of jets and the \pt of the resonance at the generator level.

The dominant background process originates from the production of two \PZ bosons from quark-antiquark annihilation $\PQq\PAQq\to\PZ\PZ\to4\Pell$, also referred to as $\PQq\PQq\PZ\PZ$, which is simulated at NLO in pQCD with \POWHEG. The differential production cross section as a function of the invariant mass of the two \PZ bosons $m_{\PZ\PZ}$ has been computed at NNLO in pQCD, but it is not yet available in a parton-level event generator. Therefore, the NNLO/NLO correction factor ($K$ factor) for the $\PQq\PQq\PZ\PZ$ background, ranging from 1 to 1.25, is applied to the \POWHEG sample as a function of $m_{\PZ\PZ}$~\cite{Grazzini2015407}. Additional NLO electroweak corrections depending on the initial-state quark flavor and kinematic properties are applied to the $\PQq\PQq\PZ\PZ$ background process, ranging from 0.7 to 1.

Another important background is the loop-induced production of two \PZ bosons through gluon fusion $\Pg\Pg\to\PZ\PZ\to4\Pell$, also referred to as $\Pg\Pg\PZ\PZ$, which is generated at leading order (LO) with \MCFM 7.0 \cite{MCFM,Campbell:2011bn,Campbell:2013una}. The LO-to-NNLO $K$ factor computed for $\Pg\Pg\to\PH\to\PZ\PZ$ is used to reweight the simulated samples of $\Pg\Pg\PZ\PZ$ events, as the $K$ factors for these two processes have been shown to be compatible by several theoretical calculations \cite{Bonvini:1304.3053,Melnikov:2015laa,Li:2015jva}.

The $K$ factor is computed as a function of $m_{\PZ\PZ}$ using the \textsc{HNNLO} v2 program~\cite{Catani:2007vq,Grazzini:2008tf,Grazzini:2013mca}, ranging from 2 to 2.6.

The background of $\PQq\PQq\to\PQq\PQq\PZ\PZ$ via $t$-channel, $u$-channel, or quartic interactions is also generated at LO with \MCFM 7.0. It is referred to as VBF$\PZ\PZ$ in what follows.
Samples of triboson events ($\PW\PW\PZ$, $\PW\PZ\PZ$) are simulated at NLO in pQCD using \MGvATNLO 2.6.5~\cite{Alwall:2014hca}, and the backgrounds from the production of top quarks associated with vector bosons ($\PQt\PAQt\PZ$, $\PQt\PAQt\PZ\PZ$, $\PQt\PAQt\PW\PW$) are generated at LO using \MGvATNLO.

Additional samples of $\PW\PZ$, Drell--Yan, and top quark-antiquark pair ($\PQt\PAQt$) production are used in the estimation of reducible backgrounds (described in Section~\ref{sec:zx}). The first two processes are simulated using \MGvATNLO at NLO in pQCD, whereas the latter is generated at NLO in pQCD with \POWHEG~\cite{Frixione:2007nw}.

Additionally, samples combining the Higgs boson production, interfering background processes and interference are generated at LO with \MCFM 7.0, considering both the ggF and VBF production mechanisms. These samples are used to study the interference between signal and background processes. The matrix element likelihood approach (MELA) \cite{Gao:2010qx,Bolognesi:2012mm,Anderson:2013afp,jhugen_5} is used to compute likelihoods using the topologies of events and matrix elements (MEs) of various processes. Based on this calculation, weights are computed to be applied to events in the samples to model the interference contributions.

All simulated samples are interfaced with \PYTHIA 8.230 \cite{SJOSTRAND2015159} to simulate parton showering and hadronization effects, with the CP5 tune \cite{tune2} used to model the underlying event.
The NNPDF 3.1~\cite{Ball:2014uwa} parton distribution functions (PDFs) are used in the generation of events.
The response of the CMS detector is modeled with \GEANTfour~\cite{Agostinelli2003250}.
The simulated events are reconstructed with the same algorithms as used for data.
Pileup (PU), the presence of additional $\Pp\Pp$ interactions in the same or nearby bunch crossings, is simulated with \PYTHIA and added to all events. The simulated samples are corrected such that their PU distribution matches the distribution observed in data.

\section{Event reconstruction and selection}
\label{sec:selection}

The particle-flow (PF) algorithm \cite{particleflow} is used to reconstruct and identify individual particles in the event with an optimized combination of the raw information from each subdetector. The reconstructed particles are referred to as PF candidates. The energy of photons is measured in the ECAL. Electrons are reconstructed by associating the reconstructed tracks in the inner tracker with the ECAL clusters, the latter additionally considering the bremsstrahlung photon showers geometrically compatible with the electron track. The electron energy is obtained from the combination of the track momentum at the primary vertex (PV) and the cluster energy, where the PV is taken to be the vertex corresponding to the hardest scattering in the event, evaluated using tracking information alone, as described in Ref.~\cite{CMS-TDR-15-02}. The muon momentum is measured from the track momenta in both the inner tracker and the muon detectors. The energy of charged hadrons is determined by combining tracks in the inner tracker with the associated energy deposits in both the ECAL and HCAL, with corrections to take into account the imperfect linear energy response of calorimeters. Finally, the energy of neutral hadrons is determined from the energy deposits in the ECAL and HCAL.

Electrons with transverse momenta $\pt^\Pe>7\GeV$ are reconstructed within the detector acceptance region, corresponding to a pseudorapidity range of $\abs{\eta^\Pe}<2.5$~\cite{CMS:2020uim}.
For each electron, the impact parameter significance, defined as the ratio of the three-dimensional (3D) impact parameter with respect to the PV to its uncertainty, is required to be $<4$. This ensures the suppression of nonprompt leptons from hadron decays or photon conversions.
Electrons are identified using a boosted decision tree algorithm based on the \textsc{XGBoost} framework~\cite{xgboost}, including electron isolation information. The algorithm was developed for the analysis in Ref.~\cite{CMS_HIG_19_001}, with an updated training performed for Ref.~\cite{Hayrapetyan2023}.

Muons with $\pt^\PGm>5\GeV$ are reconstructed within the detector acceptance region $\abs{\eta^\PGm}<2.4$, combining the information from the silicon tracker and the muon system~\cite{CMS:2018rym}. The same impact parameter requirements as for electrons are applied.
Cases where tracks in the tracker are matched to hits in only one or two muon planes are used to recover very low \pt muons that do not traverse the entire detector.
The muons considered in this analysis are selected from the muon track candidates by applying loose requirements on the track in the muon system and the inner tracker. For muons with $\pt^\PGm>200\GeV$, dedicated identification requirements are applied to improve the efficiency~\cite{CMS:2018rym}.

To distinguish the muon candidates from the \PZ boson decay from those from hadronic decays within jets, a requirement on the relative isolation $\mathcal{I}^{\PGm}<0.35$ is imposed, which was optimized in Ref.~\cite{CMS-PAS-HIG-15-004}. The relative isolation is defined as:
\begin{equation}
   \mathcal{I}^{\PGm} = \frac{1}{\pt^{\PGm}} \left(
   \sum_{\substack{\text{charged} \\ \text{hadrons}}}\pt
   + \max \left[\left(
   \sum_{\substack{\text{neutral} \\ \text{hadrons}}}\pt
   + \sum_{\text{photons}}\pt
   - \frac{1}{2} \sum_{\substack{\text{charged} \\ \text{hadrons}}}\pt^{\text{PU}}\right),
   \, \ 0\GeV \right]
   \right)
\end{equation}
where the sums are performed over PF candidates $i$ within a cone of $\Delta R(\PGm,i)<0.3$, with $\Delta R(\PGm,i)=\sqrt{\smash[b]{(\eta^\PGm-\eta^i)^2+(\phi^\PGm-\phi^i)^2}}$, where $\phi$ is the azimuthal angle in radians. The first sum involves charged hadrons originating from the PV, and the second term involves neutral hadrons and photons. The contribution of neutral particles from PU vertices, estimated as half the contribution of charged hadrons from PU, is subtracted. The total contribution from neutral particles is guaranteed to be non-negative by the max function.

Photons originating through final-state radiation (FSR) from the selected leptons are recovered to improve the accuracy of the \PZ boson candidate mass reconstruction. Particle-flow photons satisfying $\pt^\PGg>2\GeV$, $\abs{\eta^\PGg}<2.4$, and relative isolation $\mathcal{I}^\PGg<1.8$ are considered as FSR candidates. This selection was optimized in Ref.~\cite{CMS_2012}. Each candidate is associated to the closest lepton in the event and is required to have $\Delta R(\PGg,\Pell)/E_{\text{T},\PGg}^2<0.012/\GeV^2$ and $\Delta R(\PGg,\Pell)<0.5$, where $E_{\text{T},\PGg}$ is the transverse energy of the photon. For each lepton, only the photon with the lowest $\Delta R(\PGg,\Pell)/E_{\text{T},\PGg}^2$ is retained. The FSR candidates are excluded from the computation of the muon isolation.

Lepton reconstruction and selection efficiencies are measured with a tag-and-probe method, selecting \PZ boson decays both in data and in simulation, separately for electrons and muons in bins of \pt and $\eta$. All MC samples are corrected by the ratio of the efficiency in data to that in simulation.
The lepton energy scale and resolution are calibrated using known resonances decaying to two leptons in bins of \pt and $\eta$. Detailed descriptions of this calibration are given in Refs.~\cite{CMS:2020uim,CMS:2018rym}.

Hadronic jets are clustered from PF candidates using the infrared- and collinear-safe anti-$\kt$ algorithm~\cite{MatteoCacciari_2008} with a distance parameter of 0.4. The jet energy is computed from the vectorial sum of all PF candidates clustered inside it.
Since the reconstructed jet energy is on average 5--10\% larger than the true jet energy in simulation, particles from PU vertices are discarded from the jet composition, and residual corrections are applied. The corrections are derived from dijet, \PGg+~jets, \PZ+~jets, and multijet events~\cite{jec}.
In this analysis, hadronic jets are required to have $\pt^\text{jet}>30\GeV$ and $\abs{\eta^\text{jet}}<4.7$. Requirements are made on their energy fractions of different types of particles, and they must have $\Delta R>0.4$ with respect to all selected leptons and associated photons. For jets with $\pt^\text{jet}<50\GeV$, a dedicated identification algorithm is used to reject jets from PU~\cite{pileupid}.

The jet flavor is identified using the \textsc{DeepCSV} algorithm \cite{CMS_btag_pub,CMS-DP-2017-005,CMS-DP-2023-005}, a deep neural network based algorithm to tag heavy-flavor jets. Jets that pass the working point with a 68\% tagging efficiency for \PQb jets, a 10--20\% mistagging probability for \PQc jets, and a 1\% mistagging probability for light flavor jets are considered to be \PQb tagged. Since most jets from VBF stem from light-flavor quarks, vetoing \PQb-tagged jets suppresses backgrounds containing \PQb jets and improves the sensitivity to the VBF production process.

The described PF objects are used as input to the event selection, which targets events with at least four identified and isolated leptons originating from the PV. The event selection follows that employed in Ref.~\cite{CMS_HIG_19_001} and is summarized below.

Pairs of leptons with opposite charge and same flavor, whose invariant mass must be $12 < m_\PZ < 120\GeV$, are used to build \PZ boson candidates. Then, $\PZ\PZ$ candidates are formed by selecting two of the \PZ boson candidates. The \PZ boson candidate with invariant mass closest to the nominal \PZ boson mass $m_\PZ^\text{nom}=91.1876\GeV$~\cite{PDG} is defined as $\PZ_1$, and the other is defined as $\PZ_2$.
Three mutually exclusive final states are defined on the basis of the flavors of the four leptons in the event: $4\PGm$, $4\Pe$, and $2\Pe2\PGm$.
The $\PZ\PZ$ candidates must satisfy the following requirements:
\begin{itemize}
   \item Any two of the four leptons must have $\Delta R(\Pell_1,\Pell_2)>0.02$ to avoid reconstructed leptons overlapping.
   \item The $\PZ_1$ is required to have $m_{\PZ_1} > 40\GeV$ to select the on-shell \PZ boson.
   \item Among the four leptons, the one with the highest (second highest) \pt must have $\pt>20\,(10)\GeV$, to ensure high trigger and reconstruction efficiency.
   \item Every pair of oppositely charged leptons among the four selected leptons, regardless of flavor, must have $m_{\Pell^+\Pell^-}>4\GeV$, to suppress QCD multijet events.
   \item In the $4\PGm$ and $4\Pe$ final states, background processes containing a \PZ boson and a low-mass resonance such as \PJGy can pass the above selections by swapping one lepton between the two resonances. To avoid selecting such events, given a $\PZ\PZ$ candidate, two alternative \PZ boson candidates are built by swapping the two positively charged leptons. The alternative \PZ boson candidate with the invariant mass closer to $m_\PZ^\text{nom}$ is denoted as $\PZ_\text{a}$, the other as $\PZ_\text{b}$. The $\PZ\PZ$ candidate is discarded if $\abs{m_{\PZ_\text{a}}-m_\PZ^\text{nom}}<\abs{m_{\PZ_1}-m_\PZ^\text{nom}}$ and $m_{\PZ_\text{b}}<12\GeV$.
   \item The invariant mass of the $\PZ\PZ$ candidate is required to be greater than 70\GeV.
\end{itemize}
The signal region is composed of all events with at least one $\PZ\PZ$ candidate, potentially with additional leptons. If more than one $\PZ\PZ$ candidate satisfies the selection, the candidate with the highest value of a discriminant to distinguish signal and background processes, which will be defined in Section~\ref{sec:observables}, is selected.

\section{Observables}
\label{sec:observables}

A new resonance would appear as a peak in the distribution of the reconstructed four-lepton mass $\mreco$. The detector mass resolution at 125\GeV is around 1\GeV, reaching about 30, 60, and 90\GeV in the $4\Pe$, $2\Pe2\PGm$, and $4\PGm$ final states at a mass of 3\TeV, respectively.
This is mainly because the electron energy resolution worsens at high energies due to shower containment and ECAL calibration limitations, while the muon momentum resolution degrades more strongly due to reduced track curvature and increased sensitivity to alignment effects. 

\begin{figure}[!ht]
   \centering
   \includegraphics[width=0.5\linewidth]{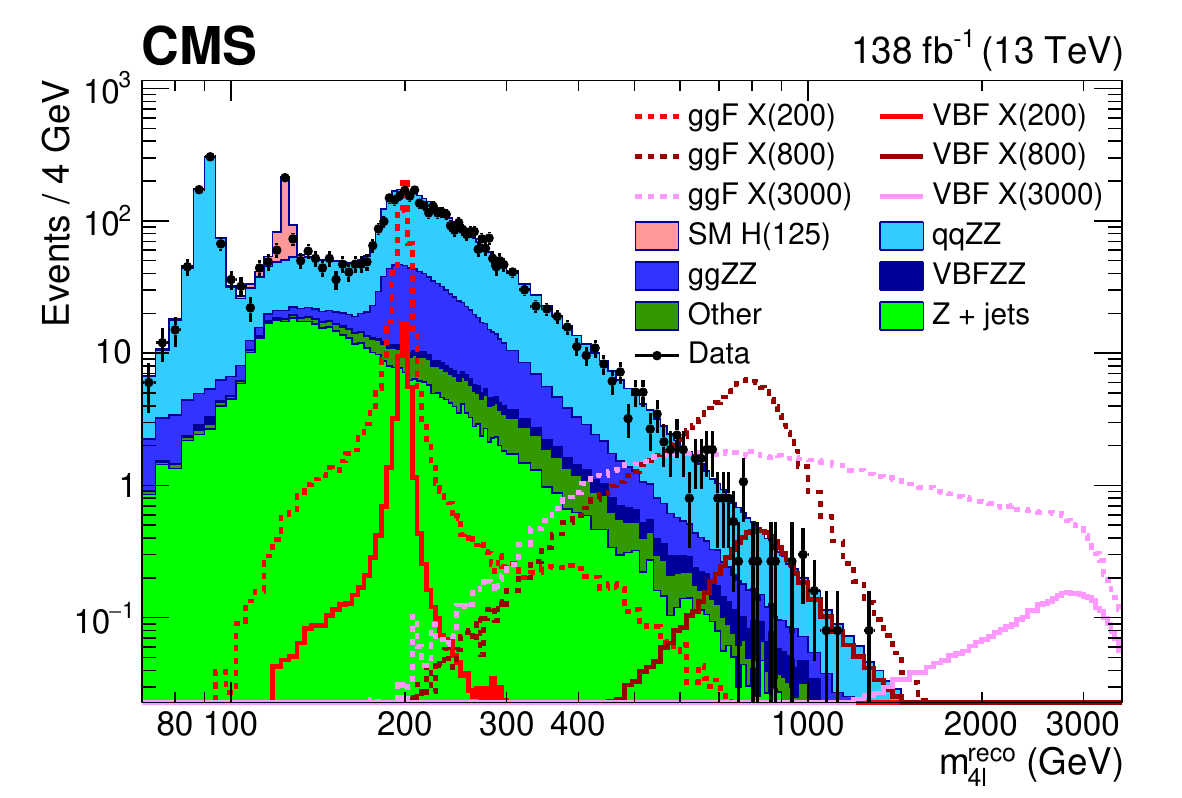}%
   \includegraphics[width=0.5\linewidth]{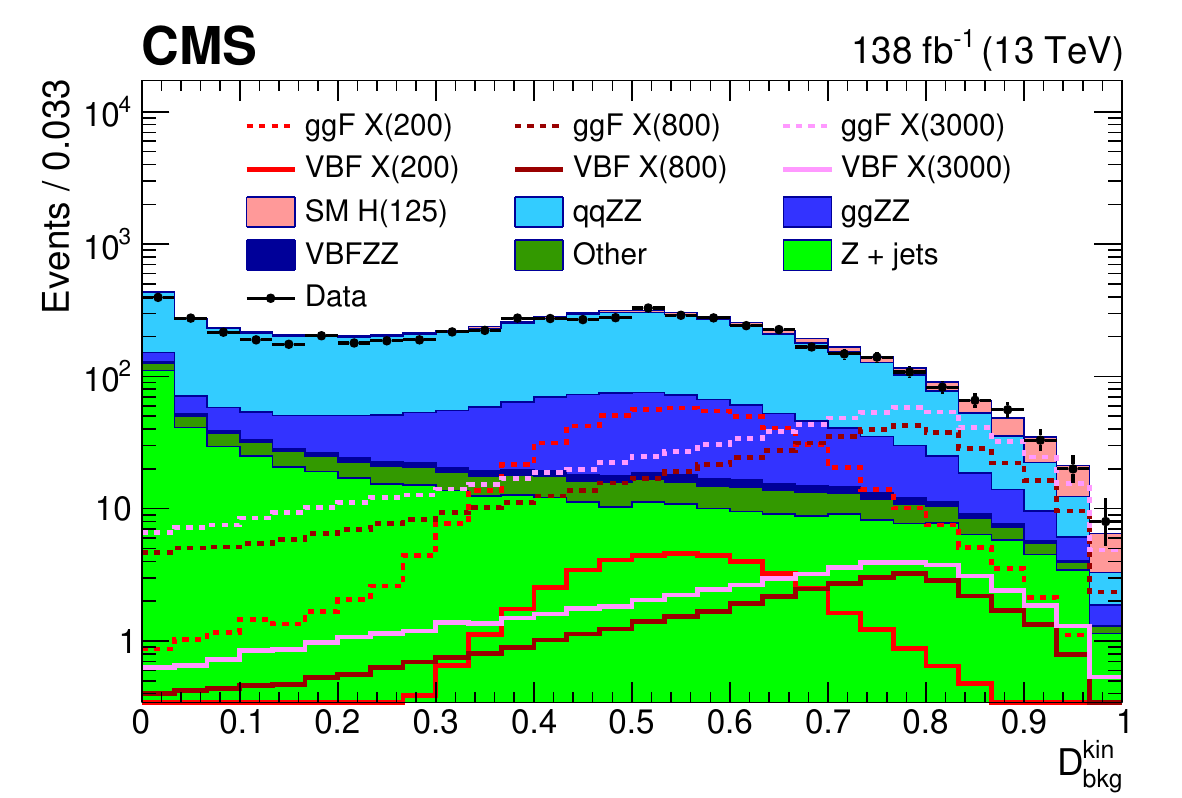}
   \caption{Left: the $\mreco$ distributions for signal and background processes estimated from the MC simulation, alongside observed data. Red and pink open histograms show the lineshapes for different signal masses. Right: The $\dbkg$ distributions for signal and background processes estimated from the MC simulation, together with the observed data. The masses of the X resonances written in the legends are in \GeV units. The widths are 1.43\GeV for $\MX=200\GeV$, 304\GeV for $\MX=800\GeV$, and 1.5\TeV for $\MX=3\TeV$~\cite{GORIA2012530}.}
   \label{fig:obs}
\end{figure}

The left panel of Fig.~\ref{fig:obs} shows the $\mreco$ distributions for various signal hypotheses, the background processes, and the observed data. All signal and background distributions are estimated from MC simulation, except the background involving jets misidentified as leptons, labelled \PZ+~jets, which is estimated from data using the method described in Section~\ref{sec:zx}. The remaining backgrounds include triboson production ($\PW\PZ\PZ, \PW\PW\PZ$) and $\PQt\PAQt$ associated with vector bosons ($\PQt\PAQt\PZ$, $\PQt\PAQt\PZ\PZ$, $\PQt\PAQt\PW\PW$), which are collectively referred to as ``other'' backgrounds.

Signal processes with different masses and produced via different production mechanisms are shown, normalized to the Higgs boson production cross section expected in the SM. The various signal processes present a narrow peak for low masses, and can be prominent above backgrounds if they have sufficiently large production cross sections. The width of the distribution increases at higher resonance masses, as expected for SM-like Higgs bosons. At 3\TeV, the ggF signal is more prominent at low $\mreco$ than the VBF signal, mostly because the gluon PDF decreases more rapidly than the quark PDF at high momentum fractions.

In addition to $\mreco$, a kinematic discriminant $\dbkg$ is considered to distinguish signal and background processes. This observable is computed using MELA, and is defined as a function of the likelihood ratios:
\begin{equation}
   \dbkg = \left [ 1 + \frac{\mathcal{P}^{\PQq\PAQq}_\text{bkg}\left ( \vec{\Omega}^{\PH\to4\Pell} | \mreco \right )}{\mathcal{P}^{\Pg\Pg}_\text{sig}\left ( \vec{\Omega}^{\PH\to4\Pell} | \mreco \right )} \right ]^{-1}.
   \label{equ:kin_discriminant}
\end{equation}
In this equation, $\mathcal{P}^{\PQq\PAQq}_\text{bkg}$ corresponds to the likelihood that the event originated from $\PQq\PAQq\to4\Pell$ and is calculated using the MEs from \MCFM; $\mathcal{P}^{\Pg\Pg}_\text{sig}$ is calculated with the MEs from \textsc{JHUGen} and defines the likelihood that the event originated from the production of a scalar resonance via ggF; and $\vec{\Omega}^{\PH\to4\Pell}$ refers to the angular phase space of the $4\Pell$ system plus the masses of the two \PZ bosons. They are computed as the probabilities of $\vec{\Omega}^{\PH\to4\Pell}$ conditional on $\mreco$.

The $\dbkg$ distributions are illustrated in Fig.~\ref{fig:obs} (right) for several signal hypotheses alongside background processes and the observed data. The backgrounds are mainly distributed around 0 and 0.5, whereas the signals are concentrated around 0.8, except for $\MX=200\GeV$, for which the distribution is concentrated around 0.5. This is because a resonance with $\MX=200\GeV$ has kinematical properties that are similar to the $\PZ\PZ$ background, which has two on-shell \PZ bosons with $\mreco$ distributed around 200\GeV. Apart from a small $\MX$ range around 200\GeV, $\dbkg$ is a powerful tool to discriminate signals at different masses against the background.

The observables $\mreco$ and $\dbkg$ are used to build two-dimensional (2D) templates, which allow us to distinguish between signal and background processes in the likelihood fit.

\section{Event categorization}
\label{sec:categorization}

To improve the sensitivity of the analysis to the VBF production process, VBF-enriched regions are defined using the topology of the hadronic jets. The criteria used to categorize events are the same as those used in Ref.~\cite{CMS_high_mass_2018}. Two mutually exclusive categories are defined: the VBF-tagged one, containing events satisfying the criteria described below, and the untagged one, containing the remaining events, which are primarily produced via ggF.

The VBF tagging criteria include the following requirements, as a VBF signal event typically contains one or two reconstructed energetic jets, which are not likely to be \PQb jets:
\begin{itemize}
    \item There is no additional electron or muon that satisfies the lepton selection discussed in Section~\ref{sec:selection} in the event.
    \item There must be at least two jets that satisfy the jet selection described in Section~\ref{sec:selection}.
    \item If there are two or three jets that satisfy the jet selection requirements, the number of \PQb-tagged jets must be $\le1$; otherwise, there must be no \PQb-tagged jets in the event. This condition is a compromise between rejecting the $\PQt\PAQt$ background which usually contains \PQb jets and retaining VBF signals that contain light jets mistagged as \PQb jets.

\end{itemize}

Requirements are also placed on the discriminant $\dtwoj$, computed with MELA and defined as:
\begin{equation}
    \dtwoj = \left [ 1 + \frac{\mathcal{P}_\text{HJJ}\left ( \vec{\Omega}^{\PH+\text{JJ}} | \mreco \right )}{\mathcal{P}_\text{VBF}\left ( \vec{\Omega}^{\PH+\text{JJ}} | \mreco \right )} \right ]^{-1},
\end{equation}
where $\mathcal{P}_\text{HJJ}$ is the probability for the event topology to be similar to that of an SM Higgs boson produced via ggF in association with two jets, and $\mathcal{P}_\text{VBF}$ is the probability for the event topology to correspond to that of an SM Higgs boson produced via VBF. The angular phase space of the $4\Pell$ system and the two most energetic jets is denoted as $\vec{\Omega}^{\PH+\text{\text{JJ}}}$. These probabilities are calculated using \textsc{JHUGen} MEs. Similar to $\dbkg$, $\dtwoj$ is defined conditional on the four-lepton mass.

\begin{figure}[!ht]
    \centering
    \includegraphics[width=0.5\linewidth]{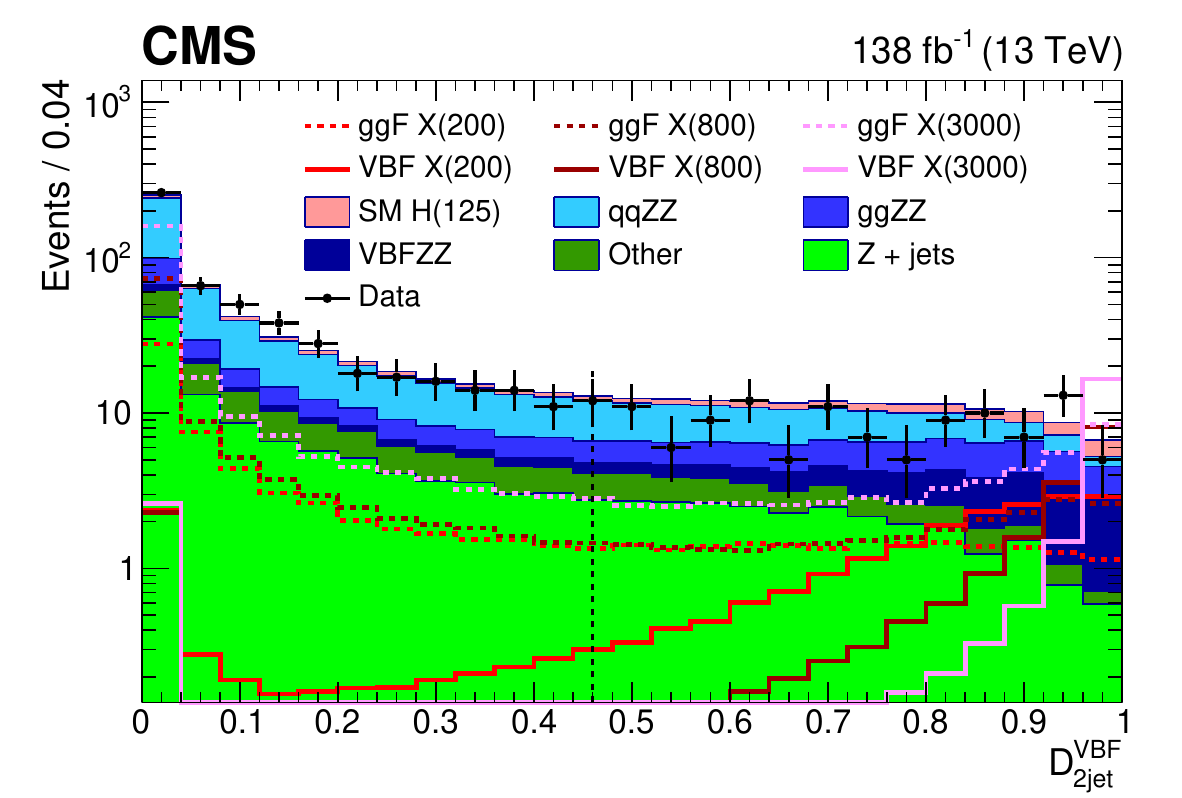}
    \caption{The $\dtwoj$ distributions for signal, background, and observed data. Only events passing the lepton and jet multiplicity requirements for the VBF-tagged category are shown. The dotted vertical line represents the threshold of $\dtwoj=0.46$.}
    \label{fig:dtwoj}
\end{figure}

Figure~\ref{fig:dtwoj} shows the $\dtwoj$ distributions of signal and background processes. Both the ggF signal and the background processes are distributed around 0, whereas the VBF signal processes are distributed around 1. For the VBF signals, a small fraction of events is distributed around 0: these correspond to the case where the two most energetic jets are not the two VBF jets but mostly ISR or FSR jets.

To be selected in the VBF-tagged category, events are required to have $\dtwoj>0.46$. This threshold maximizes the overall sensitivity to the VBF signal, taking into account the different mass hypotheses.

\section{Signal modeling}
\label{sec:signal}
To test a large number of signal hypotheses without simulating all of them, a parametric approach is used to model the different signals. The approach introduced in Ref.~\cite{CMS_high_mass_2018} is followed.

The signal model is constructed from parameterized signal templates obtained using MC simulation.
Given the mass $\MX$ and decay width $\gammaX$ of the resonance, the probability density function (pdf) of the signal process is defined as
\begin{equation}
    \mathcal{P}_\text{sig}(\mreco,\dbkg|\MX, \gammaX)=
    \{[\mathcal{M}(\mgen|\MX,\gammaX)\epsilon(\mgen)]\otimes\mathcal{R}(\mreco|\mgen)\}\mathcal{P}(\dbkg|\mreco).
    \label{equ:sigmodel}
\end{equation}
We have:
\begin{itemize}
    \item The $\otimes$ denotes a convolution.
    \item The $\mathcal{M}(\mgen|\MX,\gammaX)$ describes the theoretical lineshape of the signal, with $\mgen$ being the $4\Pell$ invariant mass at generator level.
    \item The $\epsilon(\mgen)$ is the overall signal efficiency, which accounts for the probability of the signal to fall within the detector acceptance, as well as the reconstruction, selection, and categorization efficiencies.
    \item The $\mathcal{R}(\mreco|\mgen)$ is the signal resolution function, which converts the distribution from generator level to detector level, to account for the finite energy resolution of the detector.
    \item The $\mathcal{P}(\dbkg|\mreco)$ is the template of $\dbkg$ in bins of $\mreco$, extending the formula to two dimensions.
\end{itemize}

The theoretical lineshape of the signal can be described as the differential production cross section as a function of $\mgen$, and is based on Refs.~\cite{passarino_1, GORIA2012530}. It is parameterized as
\begin{equation}
    \mathcal{M}(\mgen|\MX,\gammaX)=\sigma(\mgen)\frac{2\MX\mgen\gammaX}{[(\mgen)^2-\MX^2]^2+\MX^2\gammaX^2}.
    \label{equ:lineshape}
\end{equation}
The spin-0 propagator is described by the Breit--Wigner function; the other factors contributing to the production cross section, including the proton PDFs, the MEs, and the kinematic phase space, are incorporated as a function of $\mgen$ in $\sigma(\mgen)$, which is calculated with the \MCFM generator \cite{MCFM,Campbell:2011bn,Campbell:2013una}. Such a parameterization has been confirmed to give the same lineshape as that directly generated by \MCFM~\cite{CMS_high_mass_2018}.

\begin{figure}[!ht]
    \centering
    \includegraphics[width=0.5\linewidth]{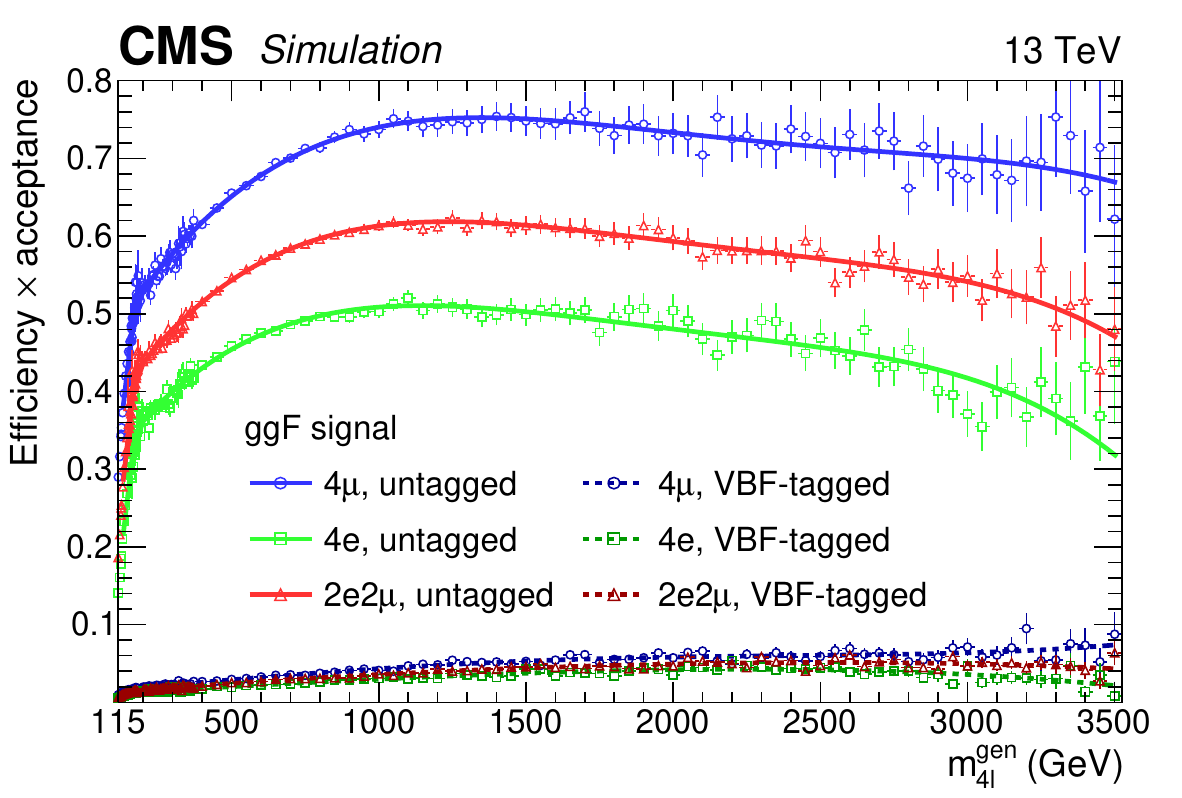}%
    \includegraphics[width=0.5\linewidth]{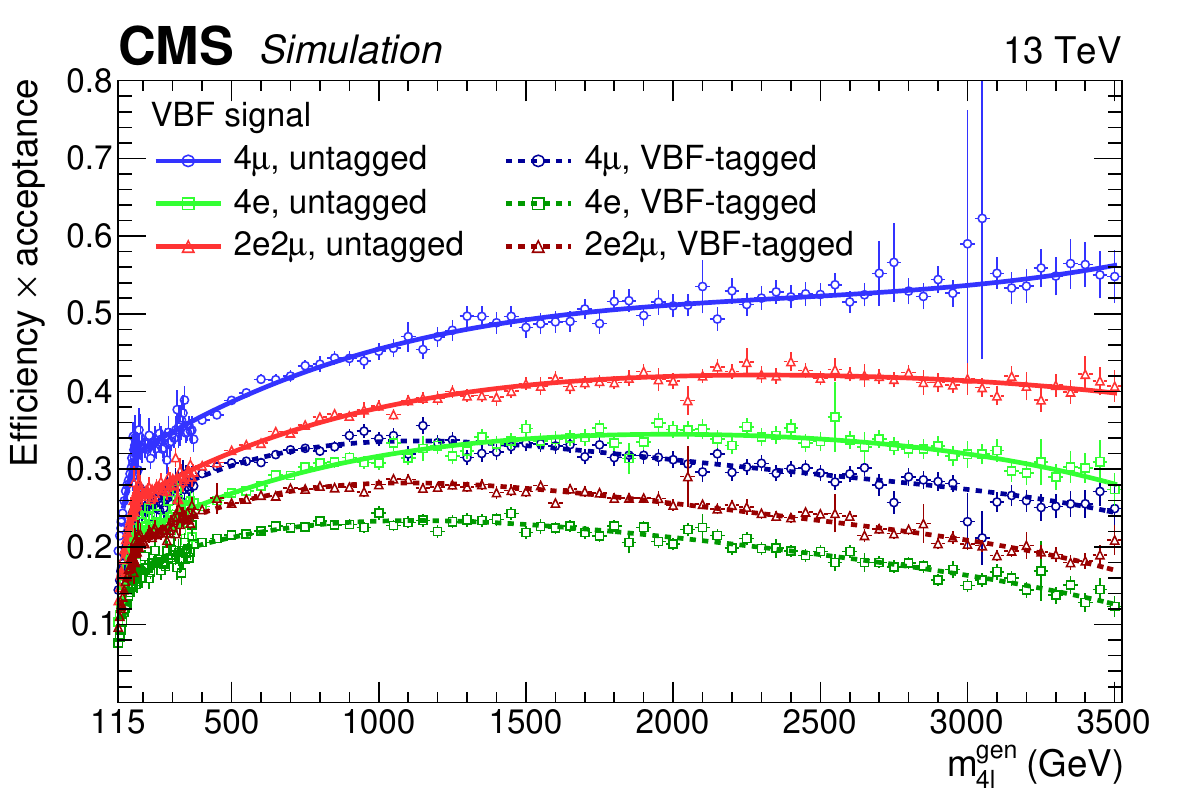}
    \caption{The product of signal efficiency and acceptance, as a function of $\mgen$, computed for the 2018 data set. The left panel shows the results for ggF signals, and the right panel shows those for VBF signals. The points are values computed from simulation, which are fitted with the corresponding curves. In each panel, the product of efficiency and acceptance for each final state and category is shown: green squares and curves represent the $4\Pe$ final state, red rectangles and curves indicate the $2\Pe2\Pgm$ final state, and blue circles and curves the $4\Pgm$ final state; the solid lines with lighter colors represent the untagged category, and the efficiencies for the VBF-tagged categories are shown in dashed lines with darker colors.}
    \label{fig:sigeff}
\end{figure}

The signal efficiency $\epsilon(\mgen)$ is defined as the number of events within the detector acceptance, satisfying the online and offline selections, and falling in a certain category, divided by the total number of generated events. The efficiency is computed for each final state, data-taking period, and production mechanism. It is estimated from MC simulation as a function of $\mgen$, and parameterized with an empirical function that is the product of a polynomial function and an error function. The $\chi^2$ divided by the number of degrees of freedom of the fits is in the range 0.6--1.3.
Figure~\ref{fig:sigeff} shows the signal efficiency evaluated from simulation for the 2018 data-taking period. In general, the efficiency is higher in the final states with muons. Signal processes have an overall efficiency of 14--80\%. More than 90\% of ggF signal events are categorized into the untagged category.
For VBF signal events, as the mass grows, the fraction of VBF-tagged events decreases. This is because the two VBF jets tend to be less energetic at higher masses, as most of the energy of the partons is needed to produce the heavy resonance. The signal efficiencies are similar for the other data-taking periods.

The signal resolution $\mathcal{R}(\mreco|\mgen)$ is described by a double Crystal Ball (DCB) function of $\Delta m_{4\Pell}=\mreco-\mgen$ defined as
{\small
\begin{equation}
\text{DCB}(\Delta m_{4\Pell}|\mu_\text{res},\sigma_\text{res},\alpha_1,n_1,\alpha_2,n_2) =
\begin{cases}
\left( \frac{n_1}{\abs{\alpha_1}} \right)^{n_1}
\re^{-\frac{\alpha_1^2}{2}}
\left( \frac{n_1}{\abs{\alpha_1}} - \abs{\alpha_1} - \frac{\Delta m_{4\Pell}-\mu_\text{res}}{\sigma_\text{res}} \right)^{-n_1},\
\frac{\Delta m_{4\Pell}-\mu_\text{res}}{\sigma_\text{res}} \le -\alpha_1& \\[8pt]

\re^{-\frac{(\Delta m_{4\Pell}-\mu_\text{res})^2}{2\sigma_\text{res}^2}},\
-\alpha_1 < \frac{\Delta m_{4\Pell}-\mu_\text{res}}{\sigma_\text{res}} < \alpha_2& \\[8pt]

\left( \frac{n_2}{\abs{\alpha_2}} \right)^{n_2}
\re^{-\frac{\alpha_2^2}{2}}
\left( \frac{n_2}{\abs{\alpha_2}} - \abs{\alpha_2} + \frac{\Delta m_{4\Pell}-\mu_\text{res}}{\sigma_\text{res}} \right)^{-n_2},\
\frac{\Delta m_{4\Pell}-\mu_\text{res}}{\sigma_\text{res}} \ge \alpha_2&
\end{cases}
\label{equ:dcb}
\end{equation}
}
where all parameters $\mu_\text{res},\sigma_\text{res},\alpha_1,n_1,\alpha_2,n_2$ are dependent on $\mgen$, with ``res'' denoting ``resolution''.

To compute the dependence on $\mgen$, all signal samples are merged with equal normalization and divided into different $\mgen$ bins. The distribution of $\Delta m_{4\Pell}$ is then extracted and fitted with the DCB function to obtain the values of the six function parameters. The dependence of each parameter $p$ on $\mgen$ is fitted with the following function:
\begin{equation}
    f_{p}(x)=a_{p,0}+a_{p,0.5}x^{0.5}+a_{p,1}x+a_{p,1.5}x^{1.5}+a_{p,2}x^2+a_{p,2.5}x^{2.5},
\end{equation}
where $x=\mgen$, $p=\mu_\text{res},\sigma_\text{res},\alpha_1,n_1,\alpha_2,n_2$. Several functional forms were studied, including polynomial functions and functions with the same form but a different order. Among these functions, $f_{p}$ was found to have the best fit performance with stable convergence and without evident overfitting effects. A simultaneous fit is then performed over all $\mgen$ bins, taking the $a_{p,i}$ as the parameters to be fitted. The method is applied individually for each production mechanism, in each final state, and for each data-taking period.

\begin{figure}[!ht]
    \centering
    \includegraphics[width=0.5\linewidth]{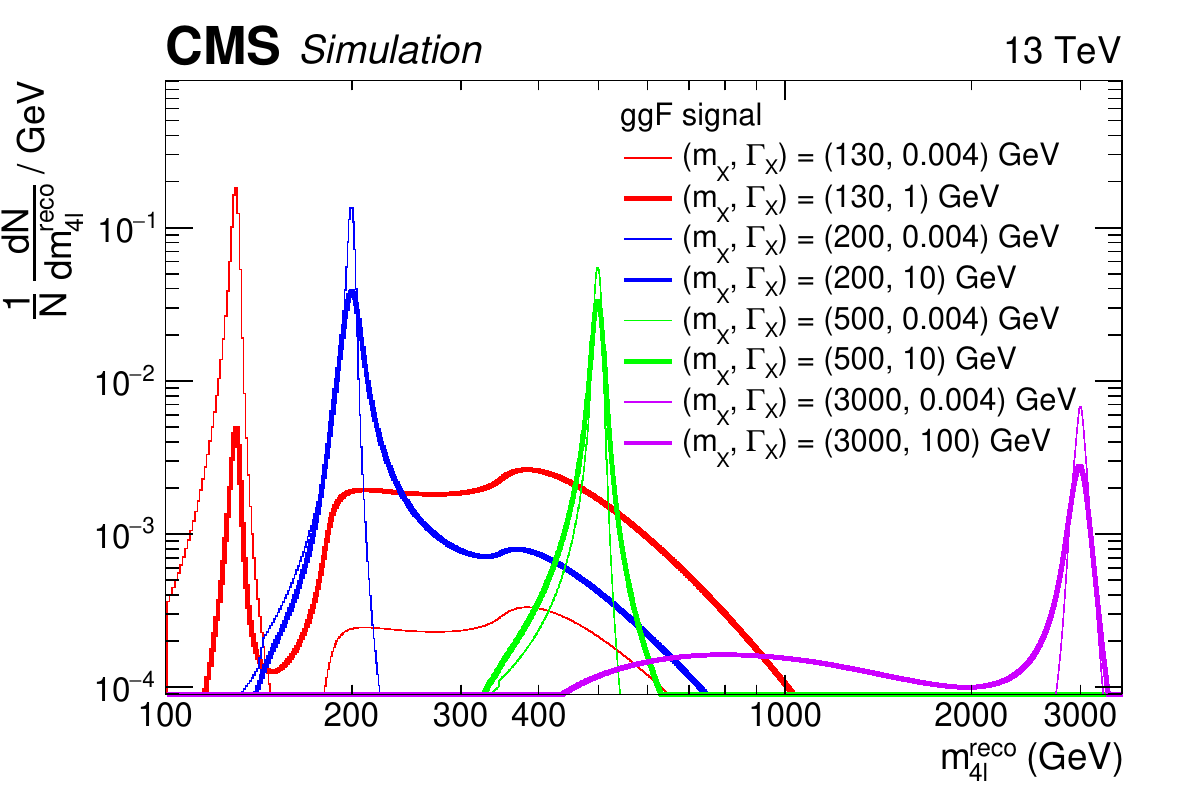}%
    \includegraphics[width=0.5\linewidth]{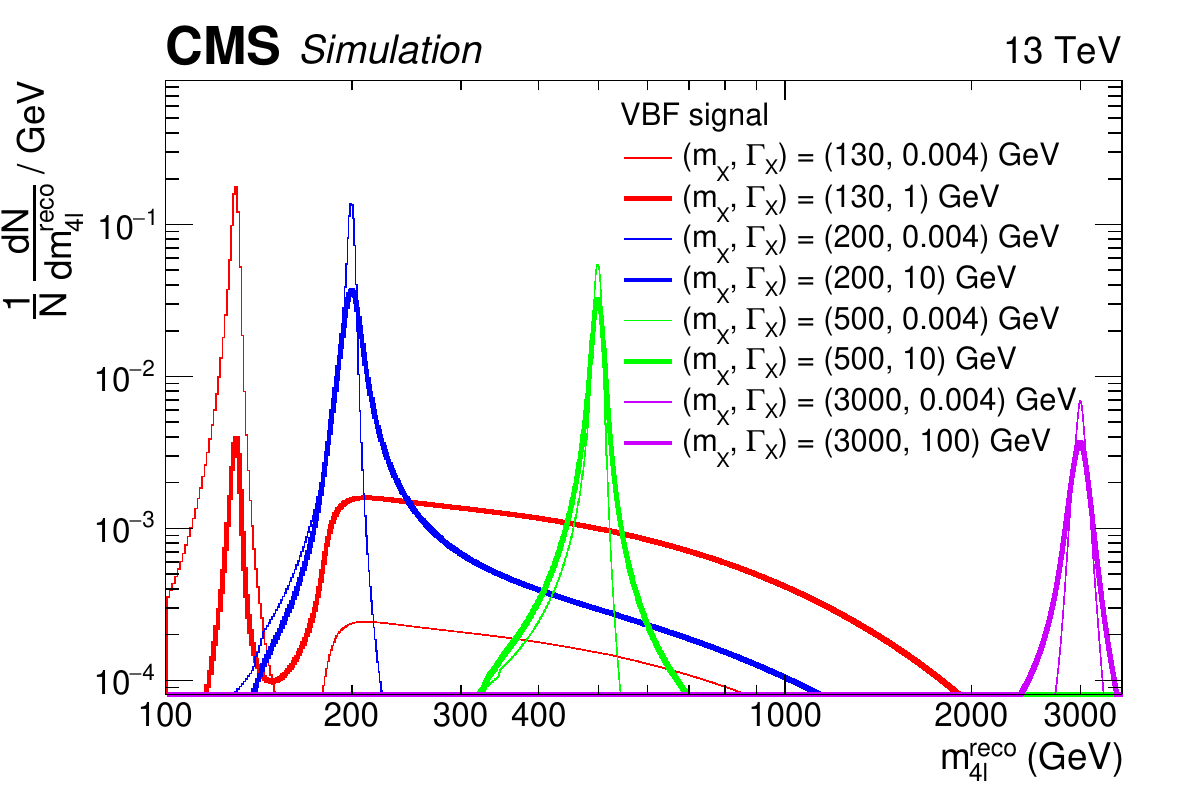}
    \caption{The $\mreco$ distributions for several values of $\MX$ and $\gammaX$ obtained from the signal model, for the ggF (left) and VBF (right) signal processes. All final states and categories are combined.}
    \label{fig:sigexample}
\end{figure}

Following the above steps, the parametric model is built as a function of $\mreco$. Examples are shown in Fig.~\ref{fig:sigexample}, with various mass and width assumptions, for the two considered signal production mechanisms. The integral of each shape is normalized to 1. For $\gammaX = 4 \MeV$, $\mreco$ is mostly distributed around $\MX$; when the width is large, a large fraction of off-shell production is expected, especially for the signals with $\MX = 130$ and 3\TeV.

\begin{figure}[!ht]
    \centering
    \includegraphics[width=0.5\linewidth]{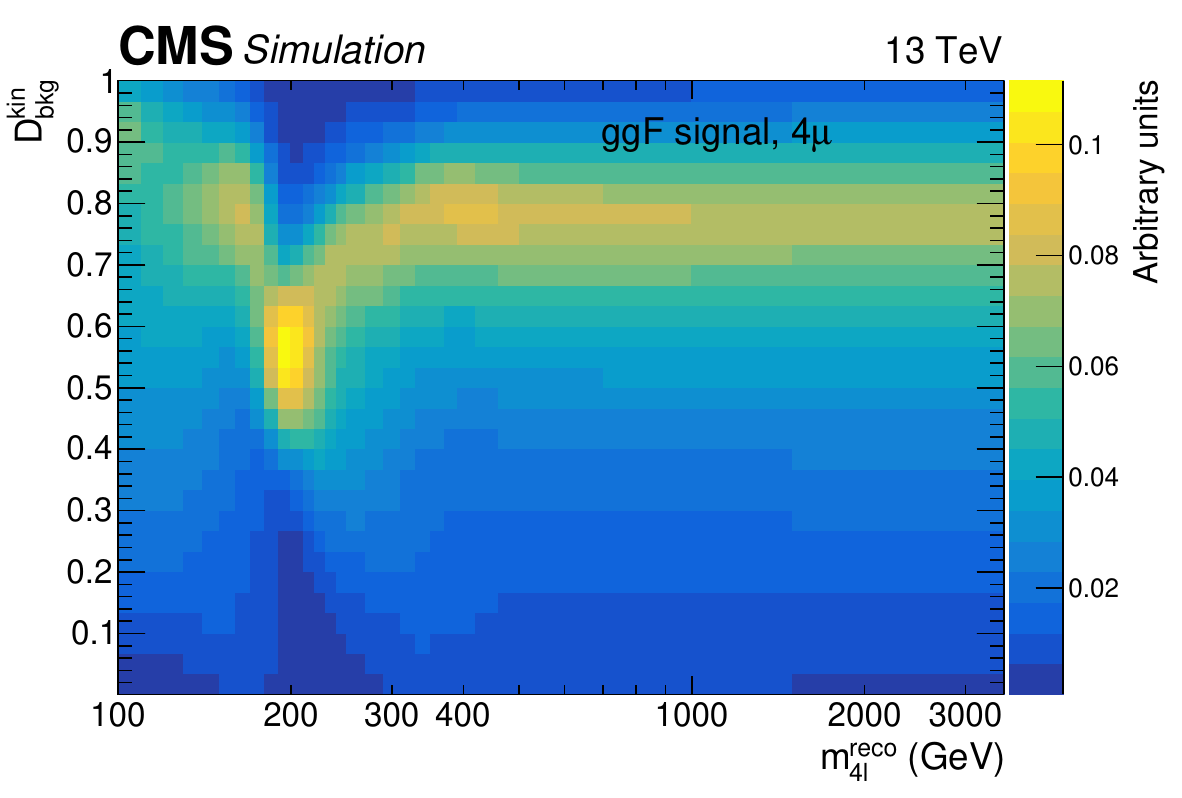}%
    \includegraphics[width=0.5\linewidth]{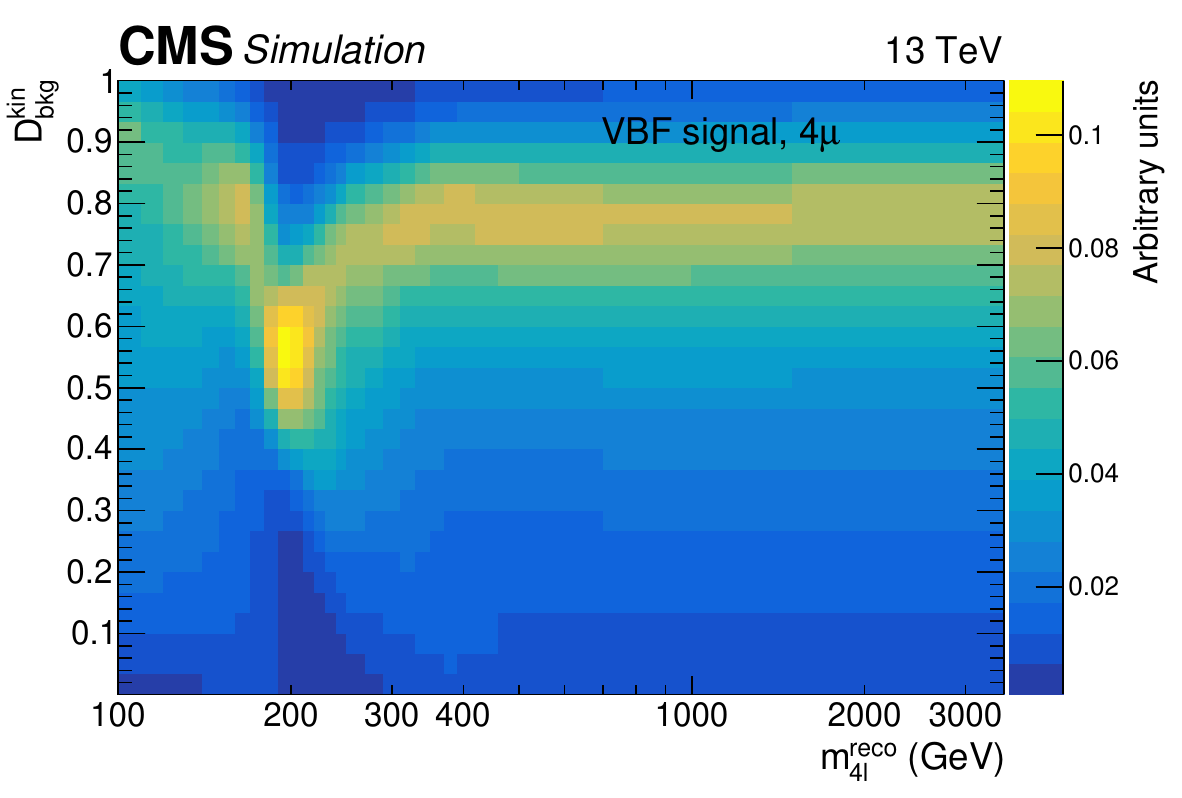}
    \caption{Expected distributions of $\mgen$ vs. $\dbkg$ for the ggF (left) and VBF (right) production mechanisms, in the $4\PGm$ final state. The distributions are estimated from the signal simulation. }
    \label{fig:2dtemp}
\end{figure}

To build the 2D pdf, the conditional pdf $\mathcal{P}(\dbkg|\mreco)$ is obtained from the $\mreco$ versus $\dbkg$ distribution extracted from the signal MC simulation. Figure~\ref{fig:2dtemp} shows examples of the distributions, for the two production mechanisms, in the $4\PGm$ final state. In each $\mreco$ bin, the distribution is normalized to unity. The distributions are similar for the two production mechanisms, since $\dbkg$ only depends on the kinematic properties of the $4\Pell$ final state.

The signal model is validated by comparing it with the $\mreco$ distribution extracted from the MC simulation. In general, they are compatible within statistical fluctuations, except in the tails of the $\mreco$ distribution for signals at $\MX>500\GeV$, which show discrepancies around 2--10\%. This is understood to be rooted in the difference between \MCFM, which generates $\sigma(\mgen)$ in Eq.~(\ref{equ:lineshape}), and \POWHEG, with which the simulation is performed. This is taken into account by adding an additional systematic uncertainty in the signal shapes.

\section{Background modeling}
\label{sec:background}
\subsection{Irreducible backgrounds}

The dominant background contribution comes from the $\PQq\PAQq\PZ\PZ$ process, with the next-largest background arising from $\Pg\Pg\PZ\PZ$ production. Another irreducible background is VBF$\PZ\PZ$. Other backgrounds, including triboson production ($\PW\PZ\PZ, \PW\PW\PZ$) and $\PQt\PAQt$ plus vector boson production ($\PQt\PAQt\PZ$, $\PQt\PAQt\PZ\PZ$, $\PQt\PAQt\PW\PW$), are also taken into account. All these backgrounds are estimated from simulation, with parameterized $\mreco$ distributions to ensure consistency with the signal model, and to smooth statistical fluctuations at high mass.

The pdfs for different backgrounds are built with different methods. For $\PQq\PAQq\PZ\PZ$ and other backgrounds, the pdf is defined as
\begin{equation}
\mathcal{P}_\text{bkg}(\mreco,\dbkg)=\mathcal{P}(\mreco)\mathcal{P}(\dbkg|\mreco),
\label{equ:backgroundpdf}
\end{equation}
where the one-dimensional pdfs $\mathcal{P}(\mreco)$ are empirical functions obtained from the simulated $\mreco$ distributions, and the conditional pdfs $\mathcal{P}(\dbkg|\mreco)$ are obtained from the $\mreco$ versus $\dbkg$ distributions in simulation.

For $\Pg\Pg\PZ\PZ$ and VBF$\PZ\PZ$, the one-dimensional pdf is built using $\mgen$ instead of $\mreco$ because, as backgrounds that interfere with signals, their generator-level lineshapes are required to derive the interference terms introduced in Section~\ref{sec:interference}. The pdf is convolved with the energy resolution function introduced in Section~\ref{sec:signal} to build the $\mreco$ pdf. It is then multiplied by the conditional pdf $\mathcal{P}(\dbkg|\mreco)$ extracted from MC simulation.

The production of a 125\GeV Higgs boson is also an irreducible background to this analysis; this process is modeled using the same approach as for the signals, by setting $\MX=125\GeV$ and $\gammaX=4.1\MeV$ in Eq.~(\ref{equ:lineshape}).

The expected yields of all the backgrounds described above are estimated from simulation.

\subsection{Reducible backgrounds}
\label{sec:zx}
The reducible backgrounds involve at least one selected lepton originating from the misidentification of a light jet, or from nonprompt decays of mesons or heavy-flavor jets. They are dominated by the Drell--Yan process, and also arise from other processes including $\PQt\PAQt$, $\PW\PZ$, $\PW\PW$, and QCD multijet events. These background processes are labelled \PZ+~jets.
The \PZ+~jets contribution is estimated from dedicated control regions in data, using the misidentification rate method of Refs.~\cite{CMS_HIG_19_001,Hayrapetyan2023}. The \PZ+~jets yield in the signal region is evaluated by multiplying the yields extracted in the control regions by the estimated lepton misidentification rates, denoted as $f_\Pe$ and $f_\PGm$. These rates are defined as the fraction of leptons satisfying all the lepton selection criteria in Section~\ref{sec:selection} except the identification and isolation requirements, to that of leptons also satisfying the identification and isolation requirements. The values of $f_\Pe$ and $f_\PGm$ are measured in another control region containing events with one \PZ candidate composed of two leptons satisfying all the selection criteria, plus exactly one additional lepton (\Pe or \PGm) without the identification and isolation requirements. This region is denoted as $\PZ+\Pell$. The misidentification rates are measured as functions of $\pt^{\Pell}$, separately for the barrel and endcap regions.

Two methods are used to estimate the \PZ+~jets contribution: the opposite-sign (OS) and same-sign (SS) methods.
In the OS method, the \PZ candidate in the $\PZ+\Pell$ region is required to have $\abs{m_\PZ-m_\PZ^\text{nom}}<7\GeV$ to improve purity; the control region used to estimate the event yield requires one \PZ candidate satisfying all the selection criteria, together with two additional OS leptons, at least one of which fails the identification and isolation requirements.
In the SS method, the mass window considered in the $\PZ+\Pell$ region is not constrained, but a dedicated approach is used to estimate the contribution of photon conversions; the control region used to estimate the event yield is defined by one \PZ candidate satisfying all the selections, together with two additional SS leptons, both without the identification and isolation requirements.
The estimated yields from the two methods are compatible within uncertainties; the final yield estimate is taken as the average of the two estimates, weighted by their statistical uncertainties, for each final state and data-taking period.
The differential distributions for the variables used in the analysis are obtained from the SS method, because the OS method introduces much larger statistical fluctuations.

The pdf for the \PZ+~jets background is built similarly to Eq.~(\ref{equ:backgroundpdf}), where $\mathcal{P}(\mreco)$ is the empirical function built from the Landau distribution~\cite{Landau:1944fvs}, fitted to the $\mreco$ distribution from the SS method. The conditional probability $\mathcal{P}(\dbkg|\mreco)$ is obtained from the $\mreco$ versus $\dbkg$ distribution from the SS method.

\section{Interference modeling}
\label{sec:interference}

In this analysis, the broad-width hypothesis is considered in addition to the narrow-width assumption (NWA). Therefore, the interference of the signal with other processes needs to be considered, as it can be significant when the signal width is large.
The modeling of the interference follows the approach used in Ref.~\cite{CMS_high_mass_2018}.
For the ggF (VBF) production mode, three possible interference contributions among the ggF (VBF) signal, ggF (VBF) 125\GeV Higgs boson production, and the $\Pg\Pg\PZ\PZ$ (VBF$\PZ\PZ$) background are considered.
The $\Pg\Pg\PZ\PZ$ and VBF$\PZ\PZ$ backgrounds will be referred to as nonresonant backgrounds in what follows. Interference terms are added to the signal and background models in the final statistical analysis. In particular, the interference between the 125\GeV Higgs boson production and nonresonant backgrounds is treated as a background since it is not related to signal processes. The pdf of each source of interference is built as
\begin{multline}
    \mathcal{P}_\text{int}(\mreco,\dbkg)= \\
    [\mathcal{P}_\text{int}(\mgen|\mathcal{P}_\text{sig}(\mgen),\mathcal{P}_\text{bkg}(\mgen),\text{ME phases})\otimes\mathcal{R}(\mreco|\mgen)]\mathcal{P}(\dbkg|\mreco).
    \label{equ:intpdf}
\end{multline}
The interference lineshape $\mathcal{P}_\text{int}(\mgen|\mathcal{P}_\text{sig}(\mgen),\mathcal{P}_\text{bkg}(\mgen),\text{ME phases})$ contains contributions from both the amplitudes and phases of the two interfering processes. The amplitudes are extracted from the lineshapes of the signals, SM Higgs boson production, and nonresonant backgrounds. The phases of the scalar propagators (signals and 125\GeV Higgs boson) are computed from the kinematical properties, whereas the phases in other parts of the MEs are derived from the corresponding generators.

The resolution functions are the same as those for the signals introduced in Section~\ref{sec:signal}, which are universal for any process with four genuine leptons.

\begin{figure}[!ht]
    \centering
    \includegraphics[width=0.5\linewidth]{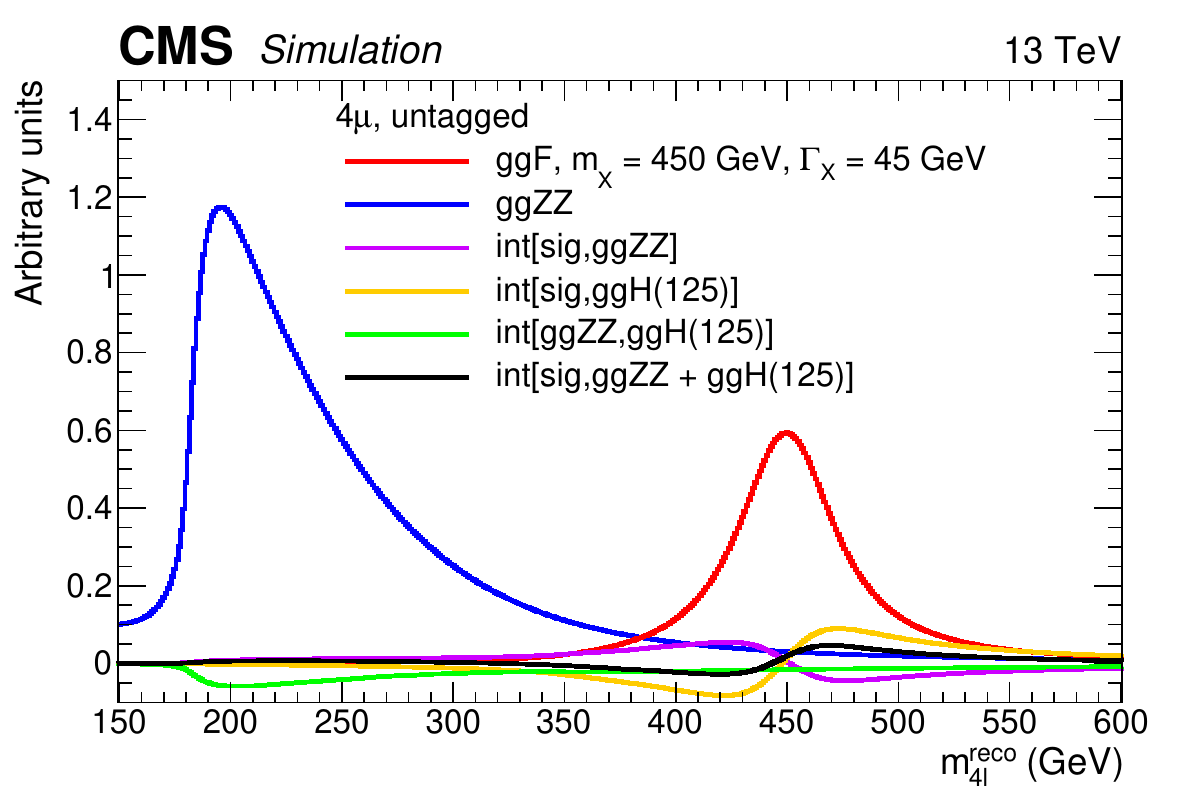}%
    \includegraphics[width=0.5\linewidth]{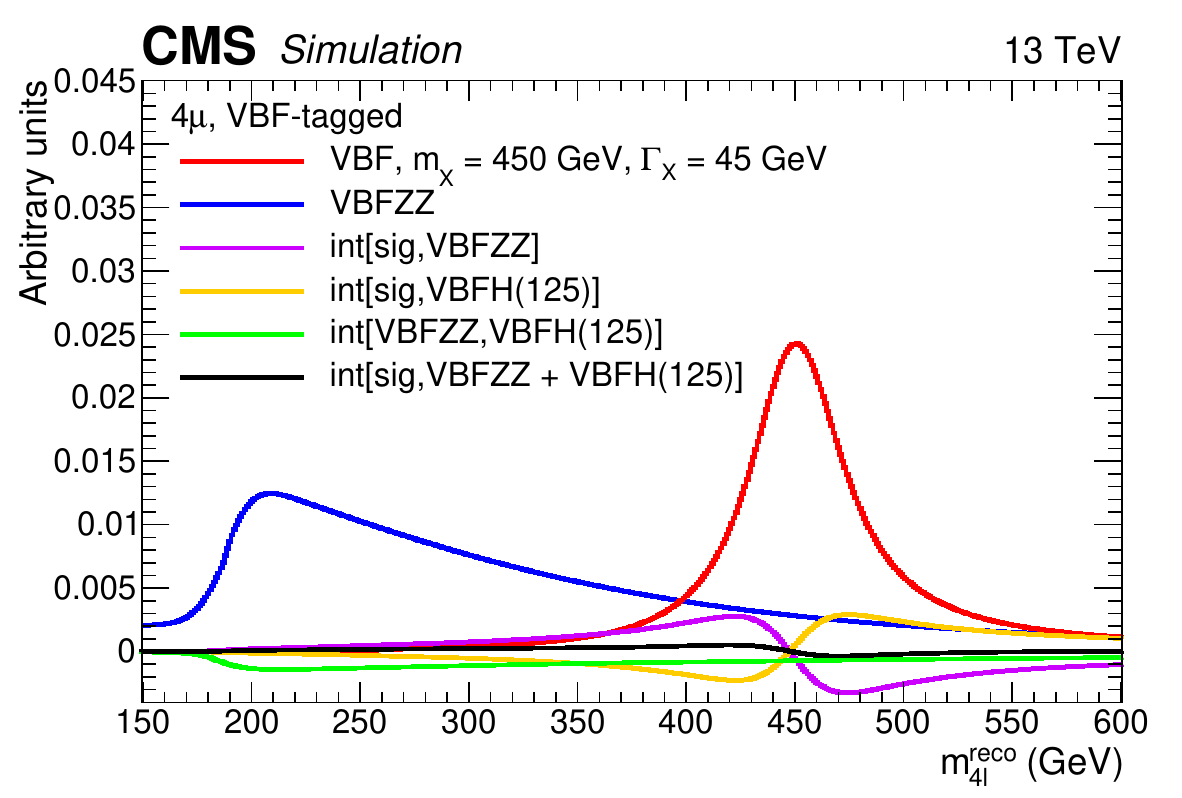}
    \caption{The left (right) plot depicts the lineshapes for the ggF (VBF) signal with $\MX=450\GeV$, $\gammaX=45\GeV$ as the red curve, the $\Pg\Pg\PZ\PZ$ (VBF$\PZ\PZ$) background as the blue curve, and interference terms as the violet, orange, and green curves. The black curve shows the interference between the signal and all other SM processes. The notation ``int[A,B]" indicates the interference between A and B. Results are shown for the $4\PGm$ final state, for the untagged category (left) and the VBF-tagged category (right). }
    \label{fig:int_lineshape}
\end{figure}
Figure~\ref{fig:int_lineshape} shows examples of interference lineshapes, alongside the corresponding signal and nonresonant backgrounds.

The backgrounds are normalized assuming their cross sections in the SM, signals are normalized such that their cross sections are $\sigma(\Pp\Pp\to\PX\to\PZ\PZ)=0.1\unit{pb}$, and the interference contributions are normalized accordingly. The interference between the 125\GeV Higgs boson and the nonresonant backgrounds is mostly negative at $\mgen>150\GeV$, as shown in Fig.~\ref{fig:int_lineshape}. The interference between the signal and the 125\GeV Higgs boson and the interference between the signal and nonresonant backgrounds have opposite signs in the region close to the signal, and almost cancel each other out in the VBF scenario.

The conditional pdf $\mathcal{P}(\dbkg|\mreco)$ for the interference terms is obtained from the $\mreco$ versus $\dbkg$ distribution extracted from MC samples reweighted using MELA variables to simulate the interference processes. Signal samples, as well as samples combining Higgs boson production, interfering backgrounds, and interference, are weighted using variables computed by MELA to estimate the interference contributions among the signal process, the 125\GeV Higgs boson, and the interfering backgrounds. The $\dbkg$ distributions for interference processes are similar to those of signal processes at a given $\mreco$.

The interference model is validated by comparing the $\mreco$ shape from the interference model to the $\mreco$ distribution obtained from MC simulation. Some discrepancies are observed for the interference between signals and nonresonant backgrounds, as well as between the 125\GeV Higgs boson and nonresonant backgrounds, stemming from the possible imperfect estimation of the phases used in Eq.~(\ref{equ:intpdf}).
Correction factors are computed as the ratios between the distribution from simulation and the interference model. These correction factors range from 1 to 1.3 and are applied to the interference model. Additional systematic uncertainties are estimated to cover the residual model-simulation difference, ranging from 9.5\% to 11\% for ggF production and 13\% to 18\% for VBF production.

\section{Systematic uncertainties}
\label{sec:systematics}

In this analysis, more than 90\% of the total uncertainty arises from statistical fluctuations in the data. However, it is still important to have a proper estimation and treatment of systematic uncertainties.

Systematic uncertainties are divided into experimental and theoretical uncertainties. Experimental uncertainties arise from imperfect knowledge of the detector and experimental measurements; theoretical uncertainties originate from the limited precision of the theoretical computations. They can affect the normalization or the shapes of each signal, background, and interference process.

\subsection{Experimental uncertainties}
The integrated luminosities are measured individually in the 2016, 2017, and 2018 data-taking periods~\cite{CMS:2016lumi, CMS:2017lumi, CMS:2018lumi}. The uncertainties range from 1\% to 3\%, and are partially correlated among the different years, leading to an overall 1.6\% uncertainty. They affect the normalization of all processes estimated from simulation, \ie, all except \PZ+~jets.

The lepton efficiency uncertainties arise from all the selection steps, and include trigger, reconstruction, and selection efficiencies. These uncertainties are dominated by the uncertainties in the lepton identification and isolation efficiency, and affect the yields of all the processes estimated from MC simulation. The lepton efficiency is measured with a tag-and-probe method, in both observed data and simulation. The same method as in Ref.~\cite{Hayrapetyan2023} is used to extract the uncertainties. Electrons and muons are treated as uncorrelated, because the selection of these objects is different. For electrons, the uncertainty ranges from 3\% to 11\% in the $4\Pe$ final state, and from 2\% to 6\% in the $2\Pe2\PGm$ final state. For muons, the uncertainty ranges from 0.2\% to 2\% in the $4\PGm$ final state, and from 0.2\% to 1.4\% in the $2\Pe2\PGm$ final state.

The uncertainties are computed individually for each process. For the ggF and VBF signals, the uncertainty values depend on the resonance masses. A parametrization is therefore built as a function of $\mgen$, and used to compute the uncertainty for any signal mass.

The lepton energy scale and resolution impact the shapes of processes convolved with the energy resolution function, including signal processes, interference contributions, and interfering backgrounds. Uncertainties for electrons and muons are considered uncorrelated. The uncertainties are estimated by comparing the $\PZ\to\Pell\Pell$ invariant mass distributions in the simulation with those from the observed data. The energy scale uncertainty is 0.15\% (0.03\%) for electrons (muons), and it affects the final states containing electrons (muons) by shifting the $\mu_\text{res}$ in Eq.~(\ref{equ:dcb}). The energy resolution uncertainty is 10\% (3\%) for electrons (muons), and it affects the final states containing electrons (muons) by modifying the $\sigma_\text{res}$ in Eq.~(\ref{equ:dcb}). The quoted values are taken from Ref.~\cite{CMS-PAS-HIG-21-019}.

The jet energy scale and resolution do not affect the total signal or background efficiency, but affect the categorization, since the VBF tagging criteria include jet selections. These uncertainties affect all processes except \PZ+~jets. Jet energy scale uncertainties originate from various sources, and they are partially correlated among data-taking periods, while the uncertainty in the jet energy resolution is considered uncorrelated among all final states and data-taking periods. In general, these uncertainties amount to around 0.1\% for the untagged category and 1\% to 4\% for the VBF-tagged category.

The jet \PQb tagging efficiency uncertainties also affect the VBF categorization efficiency, similarly to the jet energy scale and resolution. To estimate them, the \PQb tagging efficiencies are varied by their uncertainties, the number of jets that satisfy the \PQb tagging requirements is recounted, and the shifted categorization is performed. These uncertainties are estimated to be smaller than 0.1\%.

Simulated samples of events are corrected for the timing shift of the L1 trigger in 2016, and for mismodeling of PU distributions. The corresponding uncertainties affect the normalization of all processes, except \PZ+~jets. These uncertainties are typically smaller than 1\%, and are treated as uncorrelated between data-taking periods.

For the signal processes, uncertainties related to jets, PU, and the L1 timing shift effect are averaged over all mass points, since no mass dependence is observed.

The \PZ+~jets estimation uncertainties take into account the statistical fluctuations in control regions. Owing to differences in the composition between the control regions where the misidentification rates are computed and where they are applied, an additional systematic uncertainty is considered. The differences are estimated using simulation, and lead to an uncertainty of around 30\% on the yield of \PZ+~jets in the SR. The uncertainty is treated as uncorrelated among final states and data-taking periods.

\subsection{Theoretical uncertainties}

Theoretical uncertainties are correlated across all data-taking periods.

The uncertainty in the branching fraction of $\PH\to\PZ\PZ\to4\Pell$ affects the normalization of the 125\GeV Higgs boson production processes. The uncertainty amounts to around 2\%~\cite{yellow_repport_4}.

The uncertainty in the renormalization and factorization scales ($\mu_\text{R}, \mu_\text{F}$) is determined by simultaneously varying these two scales between 0.5 and 2 times their nominal value. This uncertainty affects all processes except the \PZ+~jets background, and it ranges from 1\% to 15\% depending on the process, final state, and category. For the $\PQq\PQq\PZ\PZ$ background, this uncertainty affects the $\mreco$ shape in addition to the yield.

The uncertainty in the PDFs follows the PDF4LHC recommendations, taking the root-mean-square of variations with different replicas of the default NNPDF set~\cite{Butterworth:2015oua}. The uncertainty in the strong coupling constant $\alpS$ is estimated by varying its value by 1 standard deviation (SD). The uncertainties range from 0.1\% to 7\%. For the $\PQq\PQq\PZ\PZ$ background, the uncertainty in the PDFs affects the $\mreco$ shape, just as for the $\mu_\text{R}$ and $\mu_\text{F}$ scales.

For the signal processes, the uncertainties in the renormalization and factorization scales, PDFs, and $\alpS$ are parameterized as a function of $\mgen$. For a given signal process, its lineshape is weighted by the parameterized function, and the uncertainty is computed as the difference in normalization between the weighted and nominal lineshapes.

The initial-state radiation (ISR) and FSR scales are used by \PYTHIA~\cite{SJOSTRAND2015159} to model the underlying event, as well as the fragmentation and hadronization of partons. They affect all simulated processes. The uncertainties are estimated by varying the ISR and FSR scales from 4 to 0.25 times the nominal values. The uncertainty ranges from 0.4\% to 10\%. For the signal processes, the values are averaged over all mass points.

An uncertainty is added to the signal and interference processes to account for potential differences in the lineshapes obtained from different generators. It ranges from 2\% to 10\%.
Additionally, the uncertainties accounting for the interference term mismodeling introduced in Section~\ref{sec:interference} are taken into account. They range from 9.5\% to 11\% for ggF production, and from 13\% to 18\% for VBF production.

A 10\% uncertainty in the determination of the $K$ factor used for the $\Pg\Pg\to\PZ\PZ$ prediction is applied.
For the $\PQq\PAQq\to\PZ\PZ$ background, the uncertainties affect the $\mreco$ shape, ranging from 0.1\% at low mass to 30\% at high mass.

All systematic uncertainties are treated as nuisance parameters in the fit. They are summarized in Table~\ref{tab:unc_summary}.

\begin{table}[htbp]
  \centering
  \topcaption{Summary of the experimental and theoretical uncertainties used in this analysis. Uncertainties affecting only the normalization are marked as ``norm'' in the table. Those affecting observable shapes are indicated as ``shape''.}
    \begin{tabular}{lccc}
    Source & Value & Effect & Affected processes\\
    \hline
    \multicolumn{4}{c}{Experimental uncertainties} \\
    \hline
    Integrated luminosity & 1.6\% & norm & all except \PZ+~jets\\
    $\Pe(\PGm)$ efficiency & 2--11 (0.2--2)\% & norm & all except \PZ+~jets\\
    $\Pe(\PGm)$ energy scale & 0.15 (0.03)\% & shape & mass convolution\\
    $\Pe(\PGm)$ energy resolution & 10 (3)\% & shape & mass convolution\\
    Jet energy scale & $\approx$1\% & norm & all except \PZ+~jets\\
    Jet energy resolution & $\approx$1\% & norm & all except \PZ+~jets\\
    Jet \PQb tagging efficiency & 0.1\% & norm & all except \PZ+~jets\\
    PU and L1 timing shift & 0--1\% & norm & all except \PZ+~jets\\
    \PZ+~jets estimation & 31\% & norm & \PZ+~jets \\
    \hline
    \multicolumn{4}{c}{Theoretical uncertainties} \\
    \hline
    Branching fraction of $\PH\to\PZ\PZ$ & 2\% & norm & SM \PH \\
    $\mu_\text{R}, \mu_\text{F}$ scales & 1--15\% & \begin{tabular}{@{}c@{}}$\PQq\PQq\PZ\PZ$: shape \\ others: norm\end{tabular} & all except \PZ+~jets\\
    PDF & 0.1--7\% & \begin{tabular}{@{}c@{}}$\PQq\PQq\PZ\PZ$: shape \\ others: norm\end{tabular} & all except \PZ+~jets\\
    $\alpS$ & 0.1--7\% & norm & all except \PZ+~jets\\
    ISR, FSR scales & 0.4--10\% & norm & all except \PZ+~jets\\
    Generator modeling & 2--10\% & shape & signal and interference\\
    Interference modeling & 9--18\% & norm & interference\\
    $\PQq\PQq\PZ\PZ$ $K$ factor & 0.1--30\% & shape & $\PQq\PQq\PZ\PZ$\\
    $\Pg\Pg\PZ\PZ$ $K$ factor & 10\% & norm & $\Pg\Pg\PZ\PZ$\\
    \end{tabular}
  \label{tab:unc_summary}
\end{table}

\section{Results}
\label{sec:results}

To build the complete statistical model, the signal strength is defined as
\begin{equation}
r=\frac{\sigma(\Pp\Pp\to\PX\to\PZ\PZ)}{\sigma_\text{model}(\Pp\Pp\to\PX\to\PZ\PZ)},
\end{equation}
where $\sigma(\Pp\Pp\to\PX\to\PZ\PZ)$ refers to the production cross section of the resonance multiplied by the branching fraction for $\PX\to\PZ\PZ$; $\sigma_\text{model}(\Pp\Pp\to\PX\to\PZ\PZ)$ is the reference value set in the statistical model. The fraction of the VBF signal is defined as the VBF production cross section divided by the total production cross section:
\begin{equation}
    \fvbf=\frac{\sigma(\PQq\PQq\to\PX\to\PZ\PZ)}{\sigma(\Pp\Pp\to\PX\to\PZ\PZ)}=\frac{\sigma(\PQq\PQq\to\PX\to\PZ\PZ)}{\sigma(\Pg\Pg\to\PX\to\PZ\PZ)+\sigma(\PQq\PQq\to\PX\to\PZ\PZ)},
\end{equation}
such that the signal strengths for ggF production and VBF production are
\begin{equation}
    r_\text{ggF}=r(1-\fvbf),\ r_\text{VBF}=r\fvbf.
\end{equation}
By setting $\fvbf=0$ (1), the pure ggF (VBF) production mechanism hypothesis is tested.

The sum of the pdf of each process multiplied by its predicted yield is written as:
\begin{equation}
    \mathcal{P}(\vec{x}|r,\vec{\theta})=\sum_{\text{s}=\text{ggF},\text{VBF}}[r_\text{s} S_\text{s}(\vec{\theta})\mathcal{P}_\text{s}(\vec{x}|\vec{\theta})+\sqrt{r_\text{s}} I_\text{s}(\vec{\theta})\mathcal{P}_\text{s,int}(\vec{x}|\vec{\theta})]+\sum_\text{b} B_\text{b}(\vec{\theta})\mathcal{P}_\text{b}(\vec{x}|\vec{\theta}).
\end{equation}
The first sum considers the signal and interference processes, and the second sum considers all backgrounds and the 125\GeV Higgs boson-background interference. The discriminating variables are $\vec{x}=(\mreco,\dbkg)$. The expected yields for the signal, interference, and background processes are given by $S_\text{s}$, $I_\text{s}$, and $B_\text{b}$, respectively.
The nuisance parameters $\vec{\theta}$ are used to take into account systematic uncertainties. Both the integrated yield and the differential distribution of each process can depend on the nuisance parameters.

\begin{figure}[!ht]
    \centering
    \includegraphics[width=0.5\linewidth]{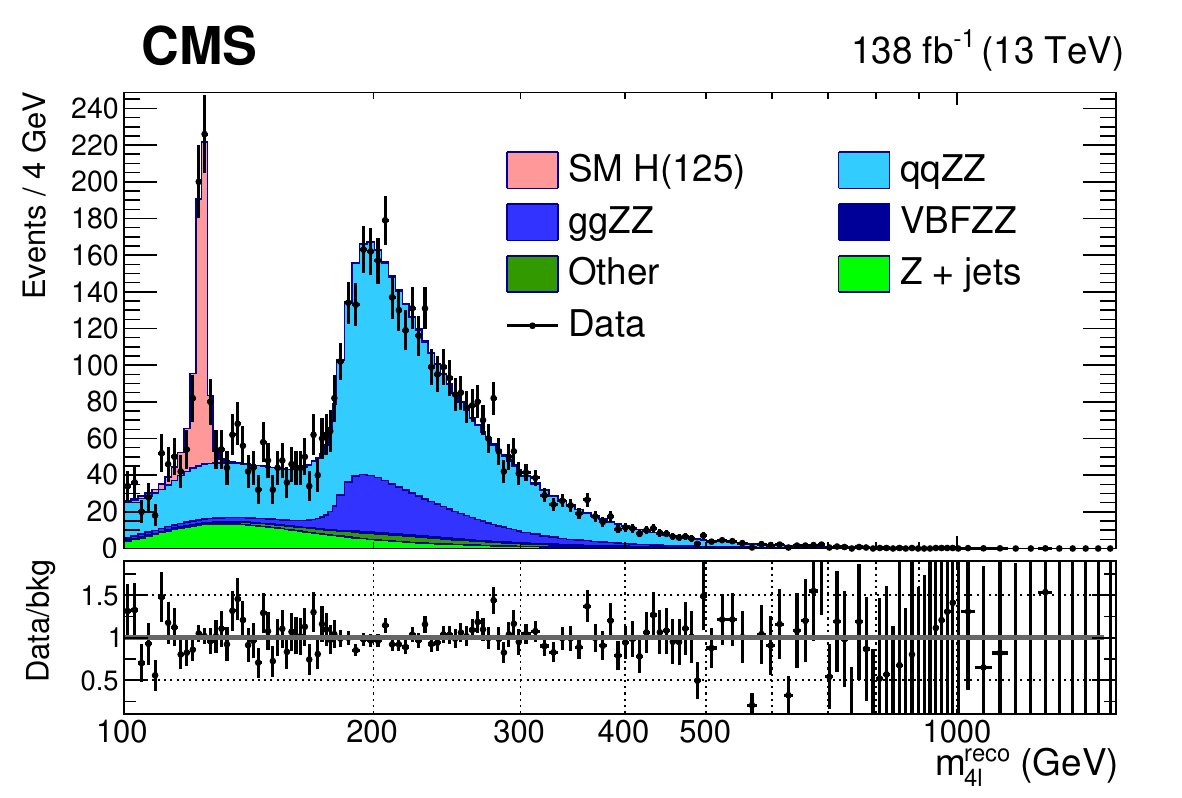}%
    \includegraphics[width=0.5\linewidth]{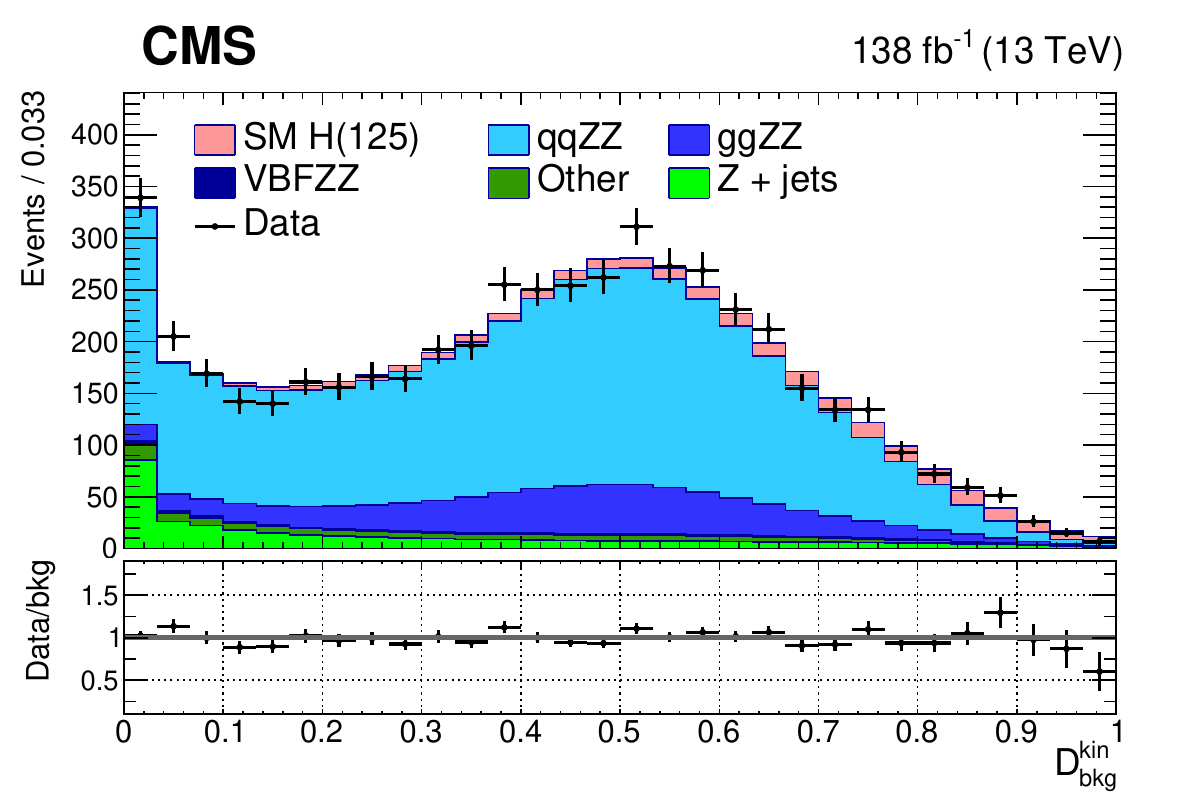}

    \includegraphics[width=0.95\linewidth]{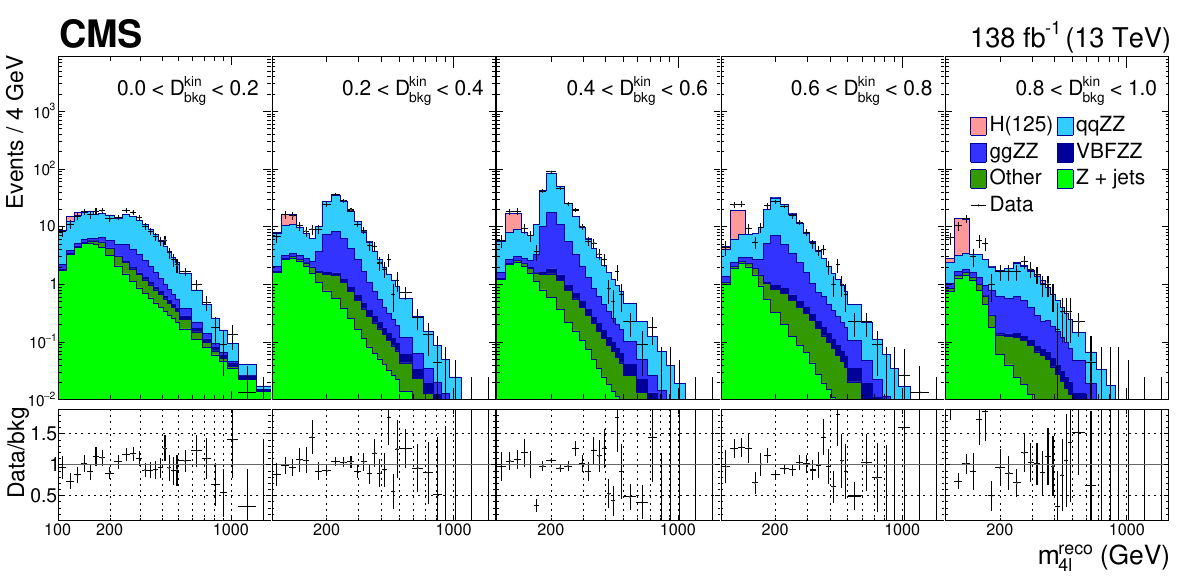}
    \caption{The $\mreco$ and $\dbkg$ distributions with the 2016--2018 data set, for backgrounds and observed data. The distributions for backgrounds are extracted from the statistical model, with all nuisance parameters at their best fit values. The upper left (right) panel shows the distribution of $\mreco$ ($\dbkg$); the lower panel shows the distribution of $\mreco$ in bins of $\dbkg$.}
    \label{fig:postfit}
\end{figure}

Extended unbinned maximum likelihood fits are performed by fitting the background-only model to the observed data, using the \textsc{Combine} tool~\cite{CAT-23-001}. Figure~\ref{fig:postfit} shows the comparison between the observed data and the background model with all parameters at their best fit values. The total $\mreco$ and $\dbkg$ distributions, as well as the $\mreco$ distributions in different ranges of $\dbkg$, are illustrated.

Upper limits at 95\% confidence level (\CL) on $\sigma(\Pp\Pp\to\PX\to\PZ\PZ)$ are computed, with all nuisance parameters floating and constrained based on the background-only fit. In the limit setting procedure, the \CLs criterion~\cite{CLS1,CLS2} is used, with the profile likelihood ratio modified for upper limits as the test statistic~\cite{CMS-NOTE-2011-005}. The asymptotic approximation~\cite{asimov} is used to estimate the test statistic distribution. For $\MX<500\GeV$, the limits computed assuming the asymptotic approximation are observed to be compatible with limits set with pseudo-experiments. At high masses, as a result of the low background yields, the limits computed with the asymptotic approximation are more stringent than those computed with pseudo-experiments. However, the differences between the two approaches are found to be within 1 SD. Given the large computational resources required to perform pseudo-experiments for all signal hypotheses, the asymptotic approximation is used to compute the limits.

\begin{figure}[!ht]
    \centering
    \includegraphics[width=0.5\linewidth]{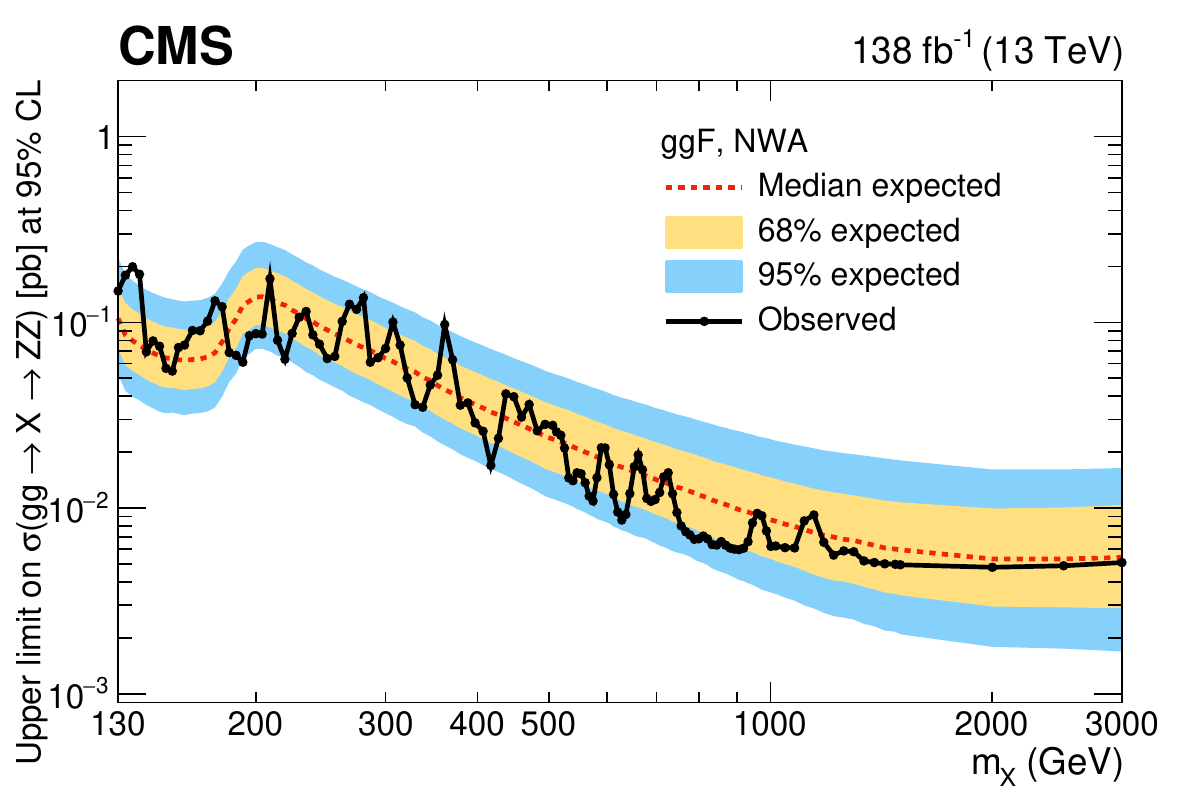}%
    \includegraphics[width=0.5\linewidth]{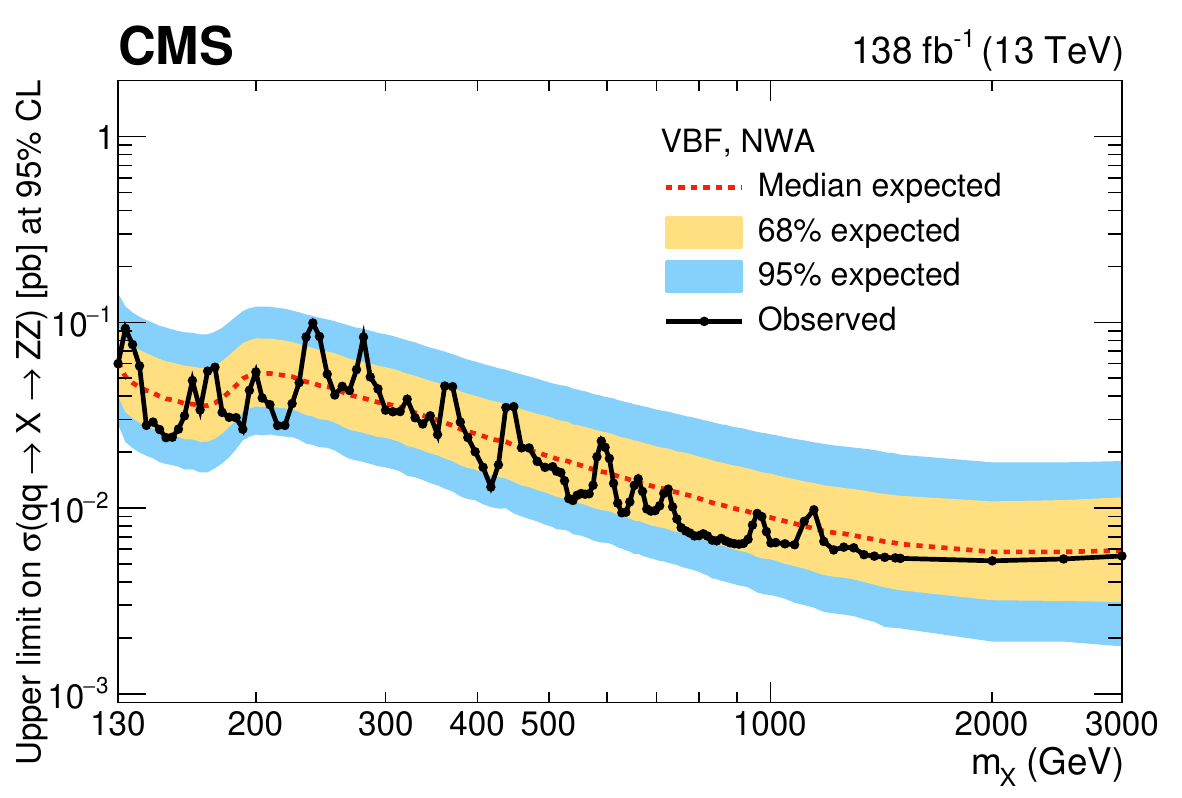}

    \includegraphics[width=0.5\linewidth]{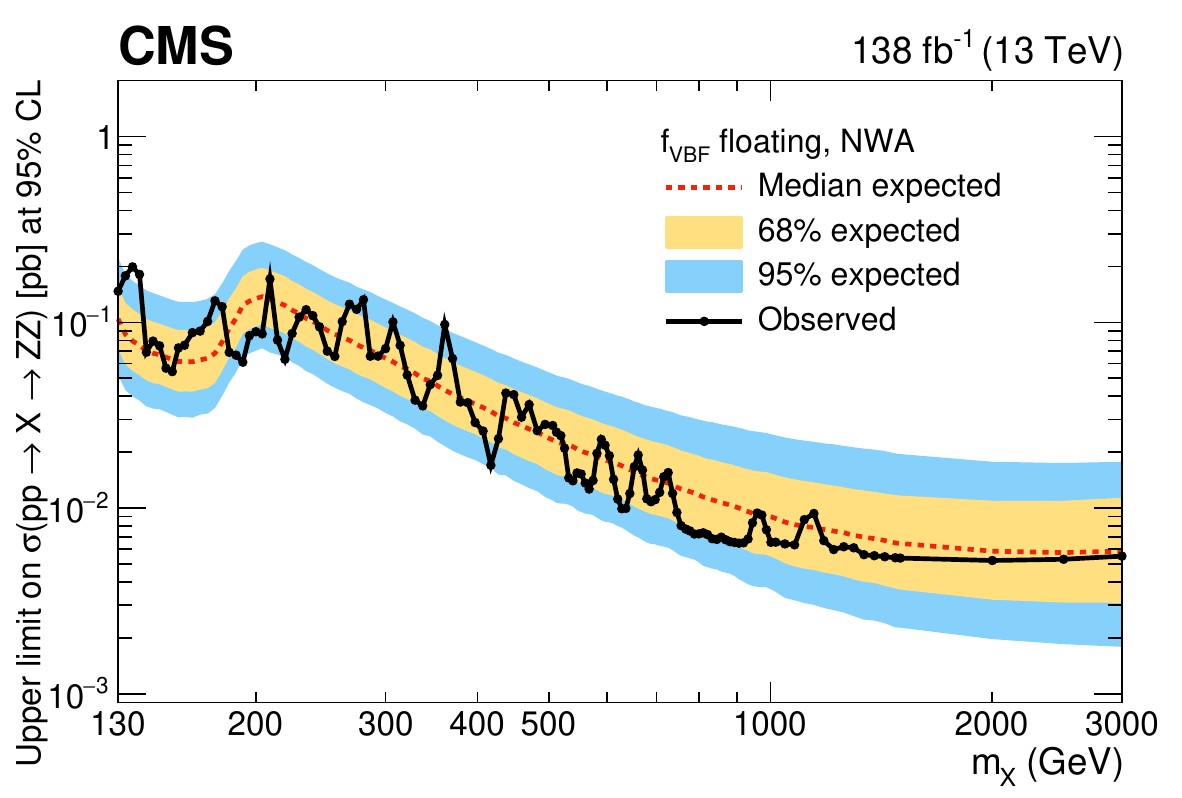}
    \caption{Observed and expected upper limits on $\sigma(\Pp\Pp\to\PX\to\PZ\PZ)$ with $\MX$ from 130\GeV to 3\TeV in the narrow-width approximation, for the ggF (upper left) and VBF (upper right) production, and with $\fvbf$ as a free parameter in the fit (lower).}
    \label{fig:nwa}
\end{figure}

Figure~\ref{fig:nwa} shows the results for the NWA, with $\gammaX$ set to 4.1\MeV, and signal-background interference excluded. The studied value of $\MX$ ranges from 130\GeV to 3\TeV. Different hypotheses for the production mechanisms are considered: pure ggF production, pure VBF production, and no assumption on the production, that is, with $\fvbf$ left as a free parameter in the fit.

The exclusion limits range from 0.05--0.1\unit{pb} in the low-mass region to 0.005\unit{pb} in the high-mass region. In general, the limits are higher at 130\GeV as a result of the 125\GeV Higgs boson background, and at 200\GeV because most of the backgrounds peak in this region. The limits for the VBF production mechanism are more stringent than the limits for ggF production, as the signal-to-background ratio in the VBF tagged category is higher than in the untagged category. The results with $\fvbf$ left as a free parameter are mostly similar to the results for pure ggF production, with the fitted value of $\fvbf$ close to 0.

\begin{figure}[!ht]
    \centering
    \includegraphics[width=0.5\textwidth]{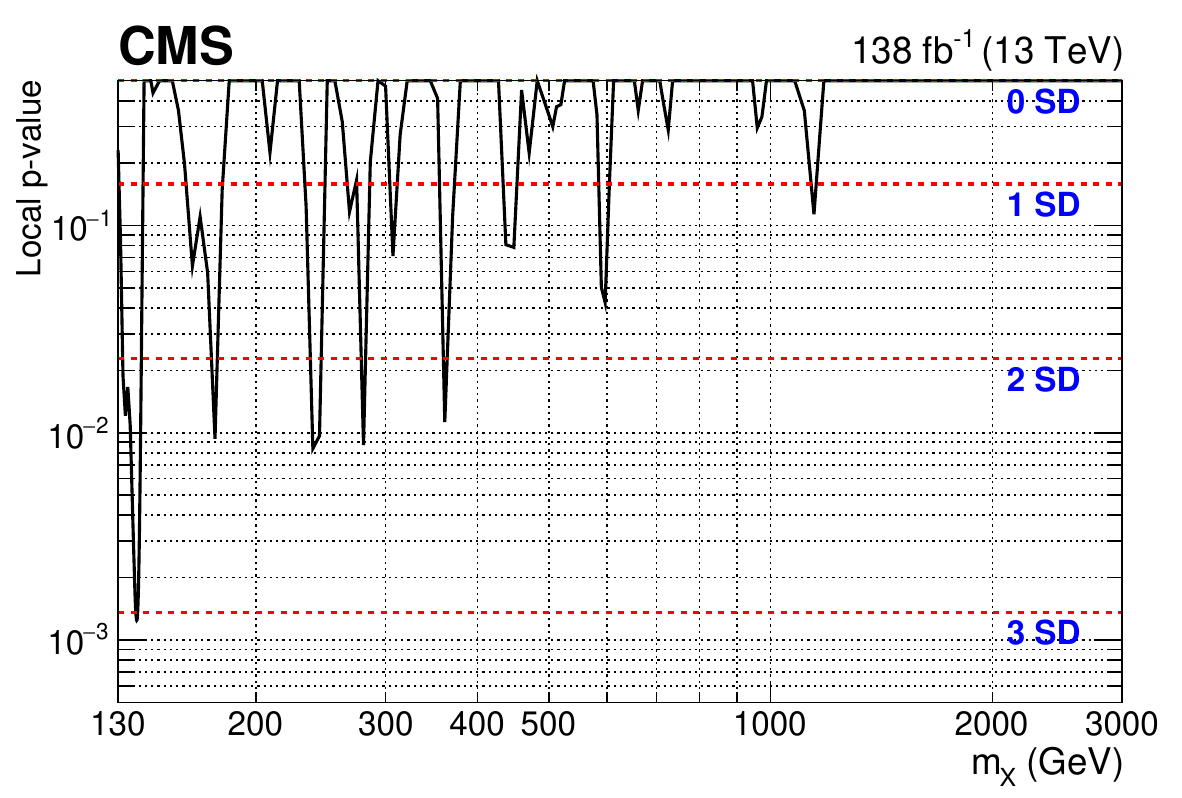}
    \caption{Local $p$-value as a function of $\MX$, with $\fvbf$ floating.}
    \label{fig:pvalue}
\end{figure}

The local $p$-value~\cite{pvalue} computed as a function of $\MX$ is shown in Fig.~\ref{fig:pvalue}. The local significance exceeds 2 SD at around 138, 176, 240, 280, and 360\GeV, and reaches a maximum of 3.0 SD at 137.8\GeV.

To consider the look-elsewhere effect~\cite{Gross:2010qma}, the global significance is computed by estimating the probability of finding such an excess in the 130--145\GeV mass region using 1000 toys. The global significance is found to be 1.8 SD.

In some previous studies, excesses around 600--700\GeV have been observed with global significances of 0.9--2.8 SD~\cite{ATLAS_high_mass_2022,00_2_gg_ATLAS,CMS_HIG_21_011}, which can be fitted by a resonance suggested by BSM theories~\cite{universe9020099,Consoli2024}. No such excess is found in this search.

\begin{figure}[!ht]
    \centering
    \includegraphics[width=0.5\linewidth]{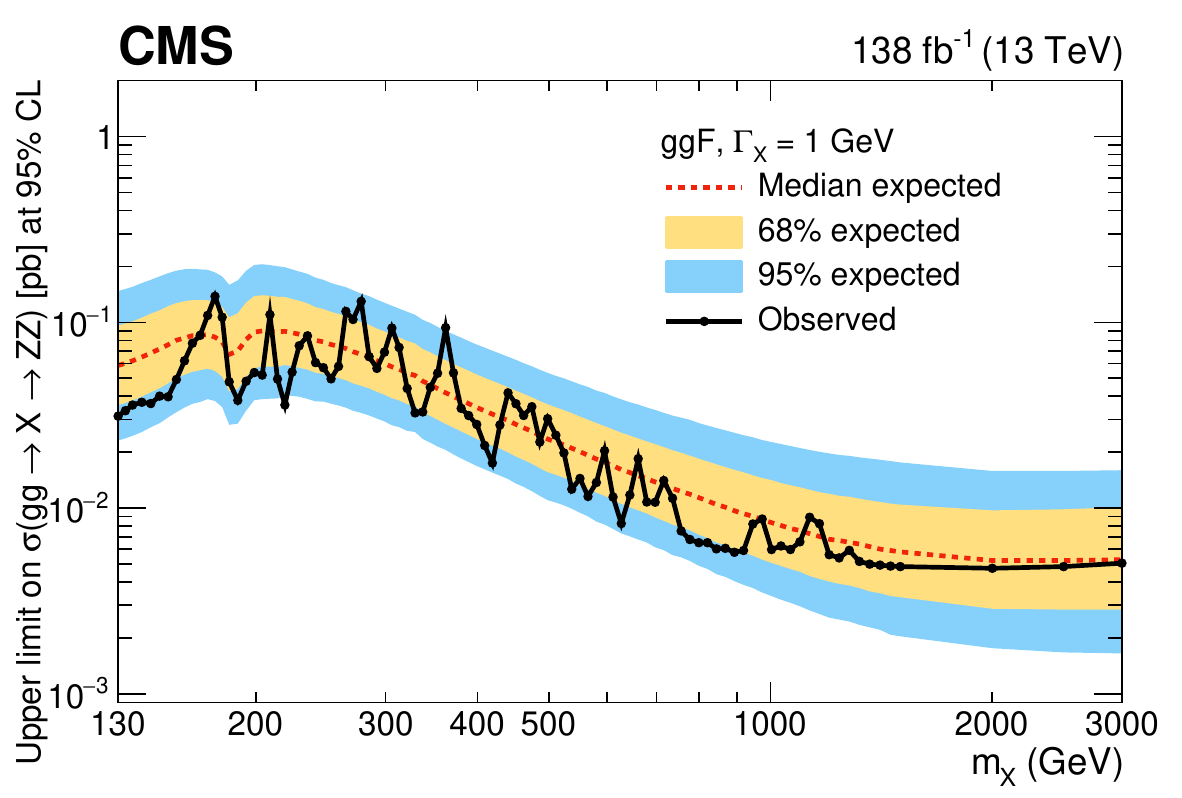}%
    \includegraphics[width=0.5\linewidth]{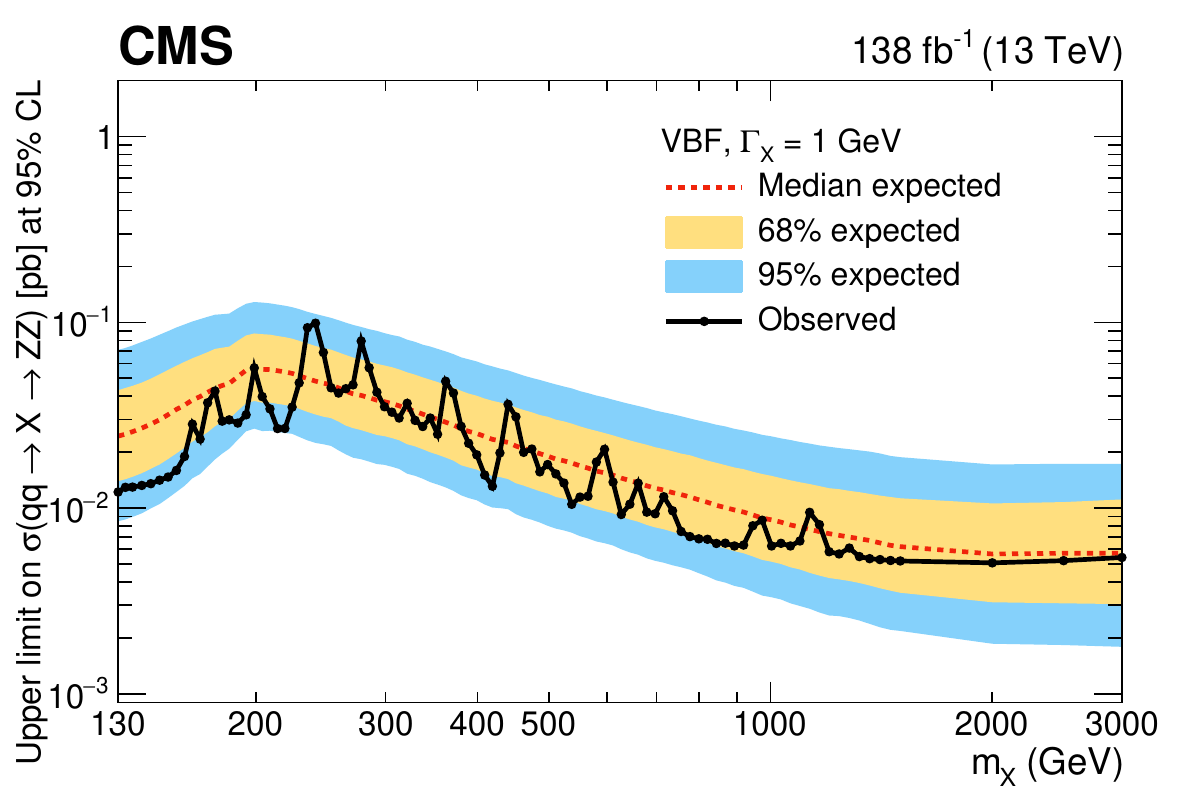}

    \includegraphics[width=0.5\linewidth]{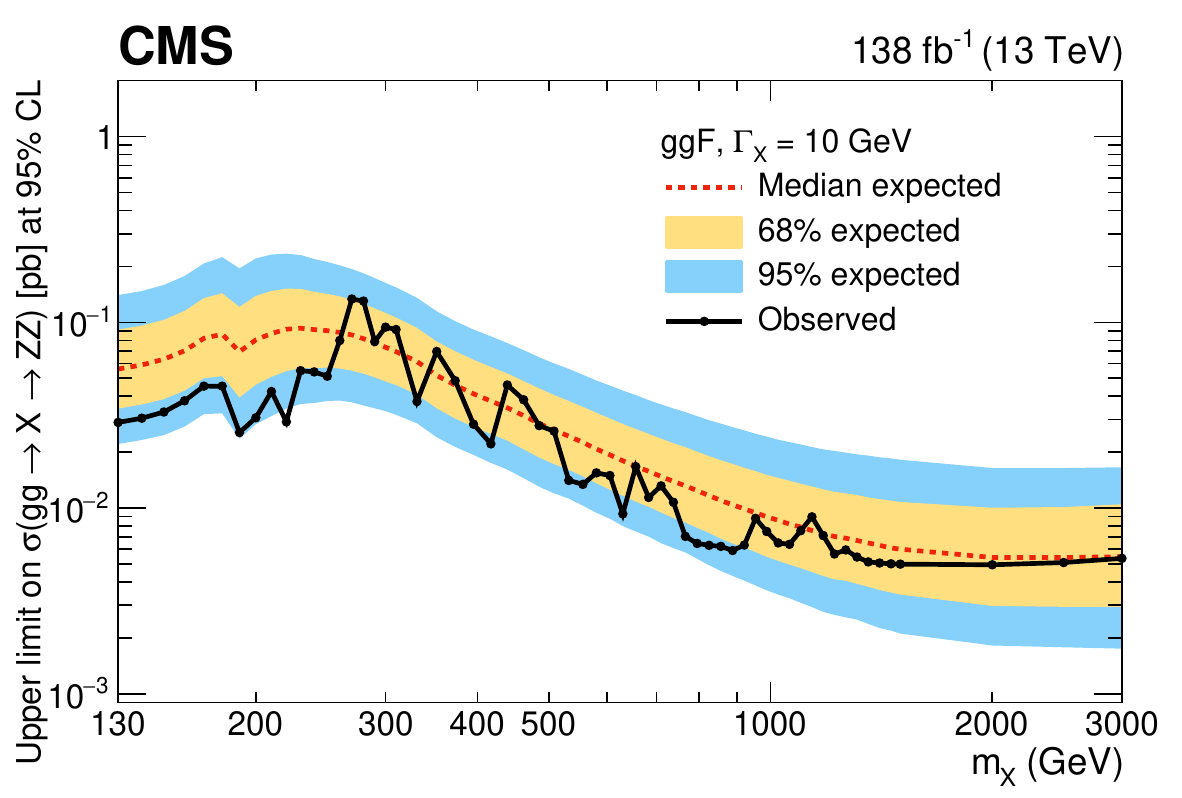}%
    \includegraphics[width=0.5\linewidth]{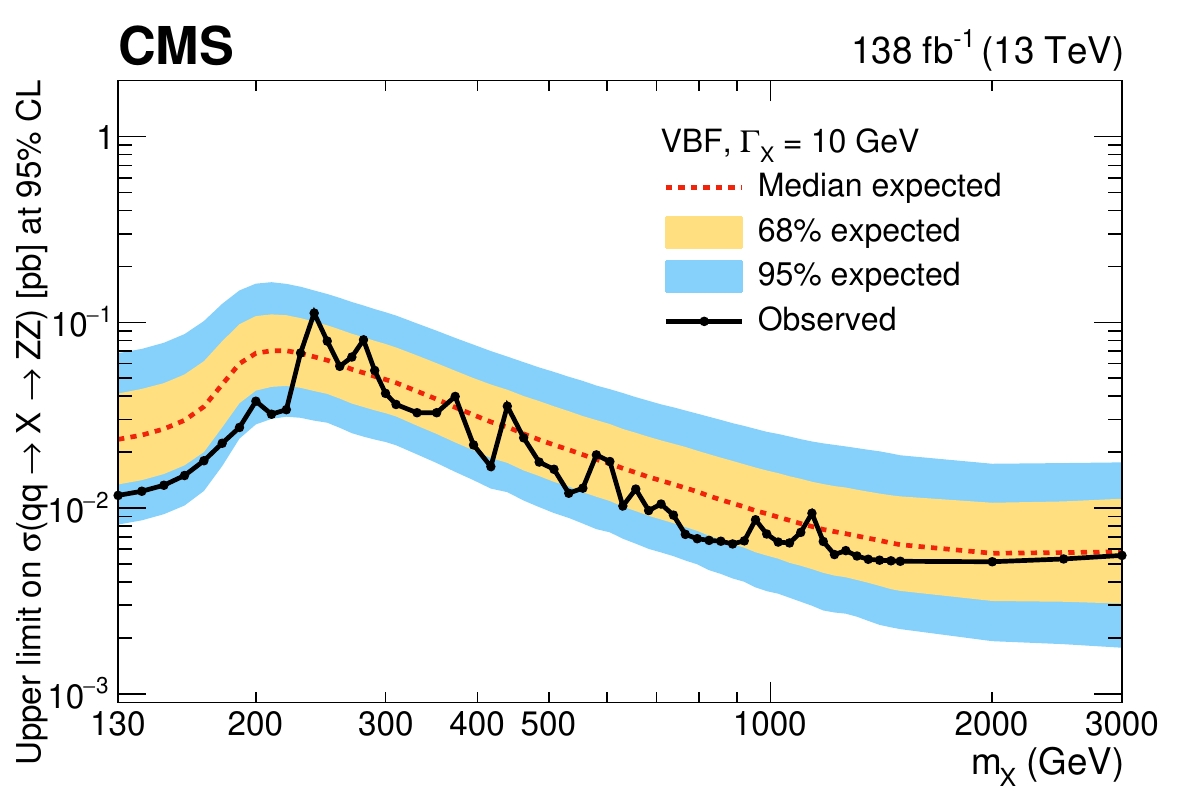}

    \includegraphics[width=0.5\linewidth]{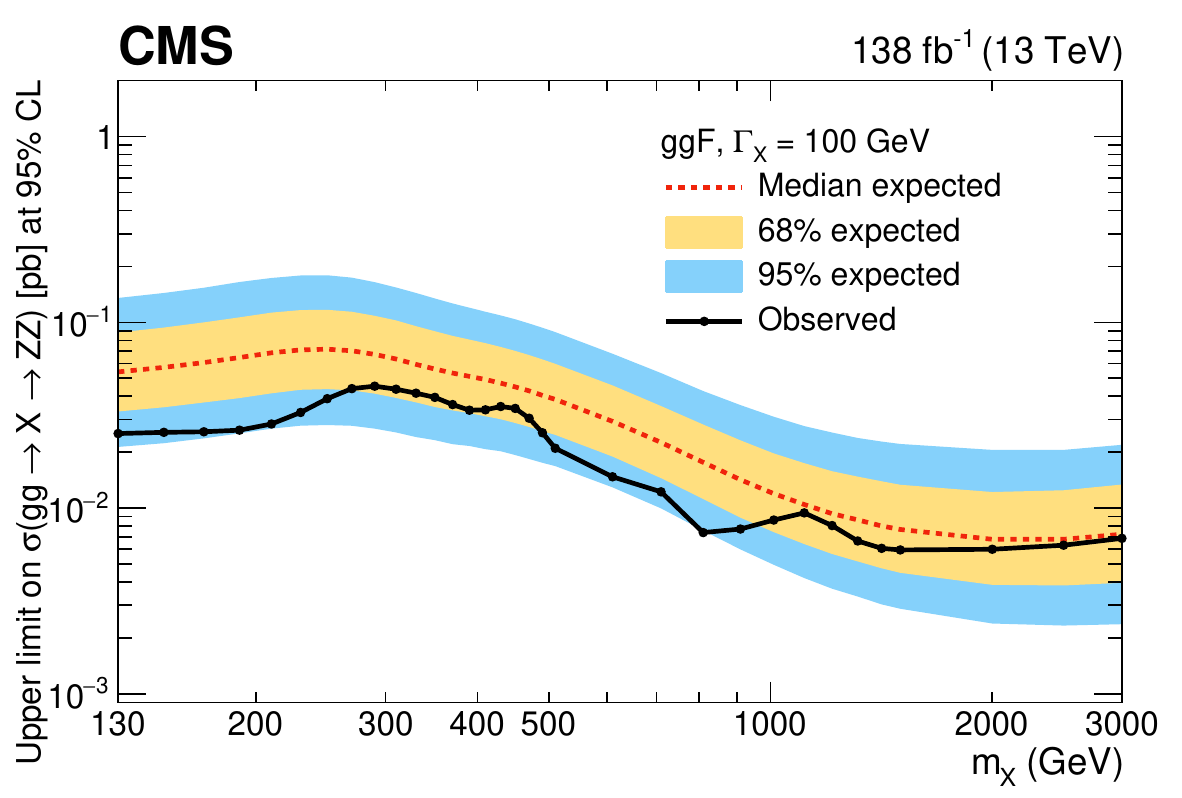}%
    \includegraphics[width=0.5\linewidth]{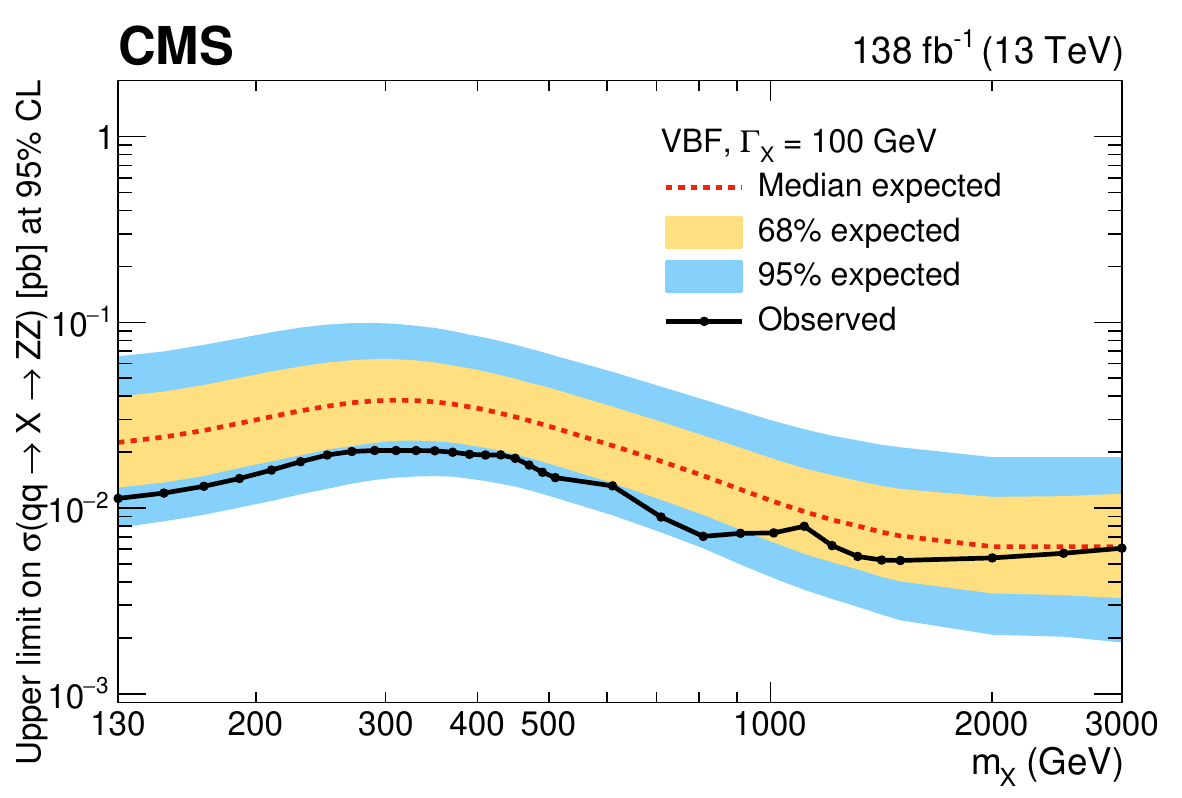}
    \caption{Observed and expected upper limits on $\sigma(\Pp\Pp\to\PX\to\PZ\PZ)$ with $\MX$ from 130\GeV to 3\TeV and $\gammaX$ equal to 1 (upper), 10 (middle), and 100 (lower) \GeV. The left column shows the results for pure ggF production and the right column shows the results for pure VBF production.}
    \label{fig:fwa}
\end{figure}

Broad-width assumptions are also tested, with $\gammaX$ fixed to 1, 10, or 100\GeV. The results are shown in Fig.~\ref{fig:fwa}. From the NWA to the 1\GeV width assumption, the limits are stronger at low masses mainly because of the off-shell production present above 200\GeV, where the signal efficiency is much higher, especially for signal masses below 200\GeV. In addition, the interference terms contribute to these stronger limits. With the decay width increasing, the limits become slightly less stringent, mainly because of the more smeared $\mreco$ distributions.

The limits for $\MX$ around 200\GeV, where the background contribution is the largest, become more stringent with $\gammaX=100\GeV$ than those with $\gammaX=10\GeV$. The reason for this is that as the width grows, the $\mreco$ distribution becomes broader and more signal events fall in the region with lower background contributions.

Comparing the observed and expected results, no significant excess is observed. At $\MX<200\GeV$ with $\gammaX=1$ and 10\GeV, as well as over the whole mass range for $\gammaX=100\GeV$, the observed results are systematically lower than the median expected results. The total yield of the background model is slightly larger than the number of observed events. The observed results are, therefore, more stringent since a larger signal strength would increase the disagreement with the observed data. This is more clearly seen for the signal hypotheses that have a broad spread in their mass distribution, because the deficit of data with respect to the background expectation is correlated over a wide range through the signal lineshape.
Signals at low masses ($\MX<200\GeV$) exhibit a broad spread mainly as a result of the presence of off-shell production as shown in Fig.~\ref{fig:sigexample}. For these low-mass signals, the off-shell production contributes 50--96\% for $\gammaX=1\GeV$, 92--98\% for $\gammaX=10\GeV$, and greater than 99\% for $\gammaX=100\GeV$. All signals with $\gammaX=100\GeV$ exhibit a broad spread in the mass distribution as a result of their large width.

\begin{figure}[!htp]
    \centering
    \includegraphics[width=0.75\linewidth]{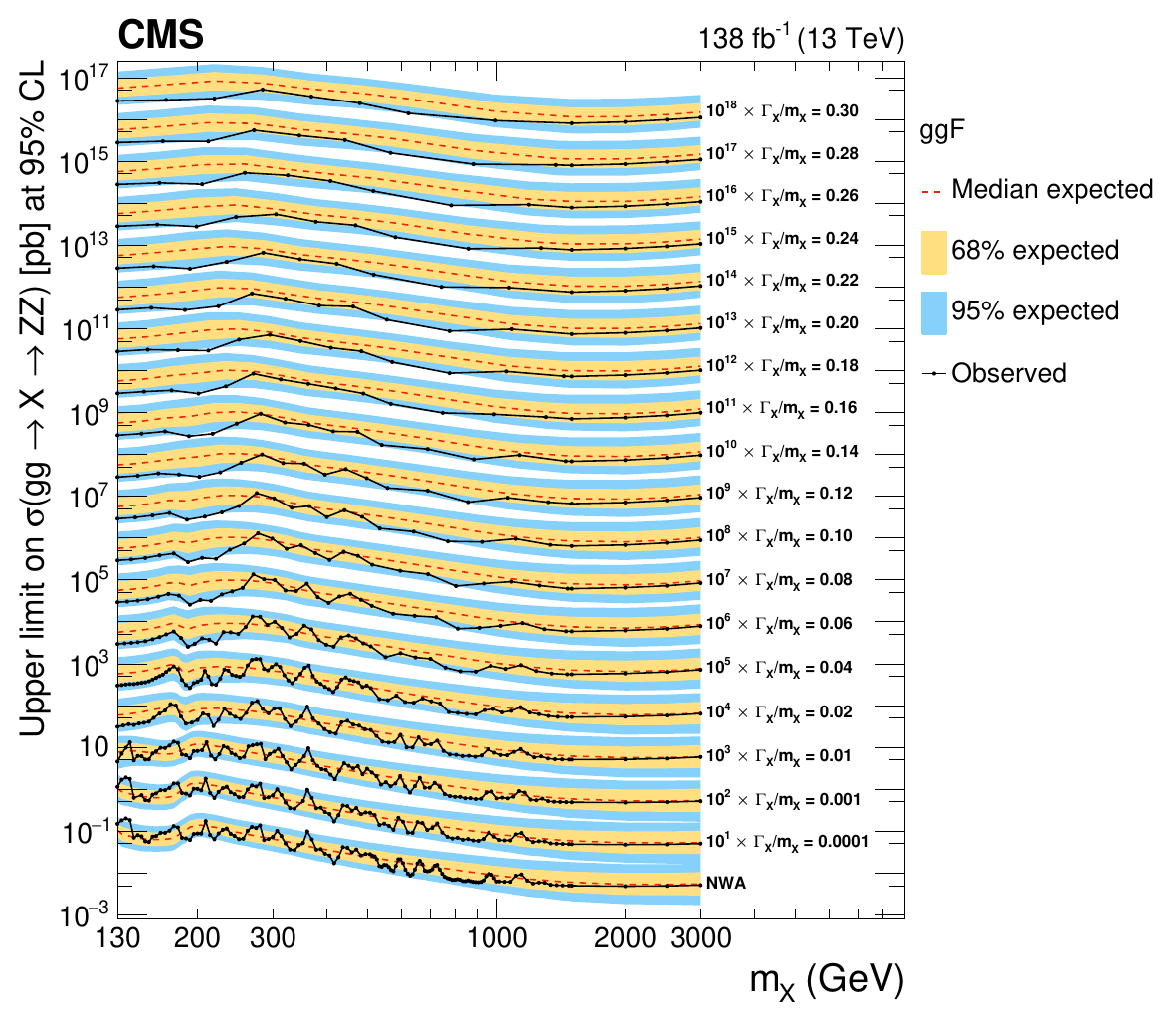}

    \includegraphics[width=0.75\linewidth]{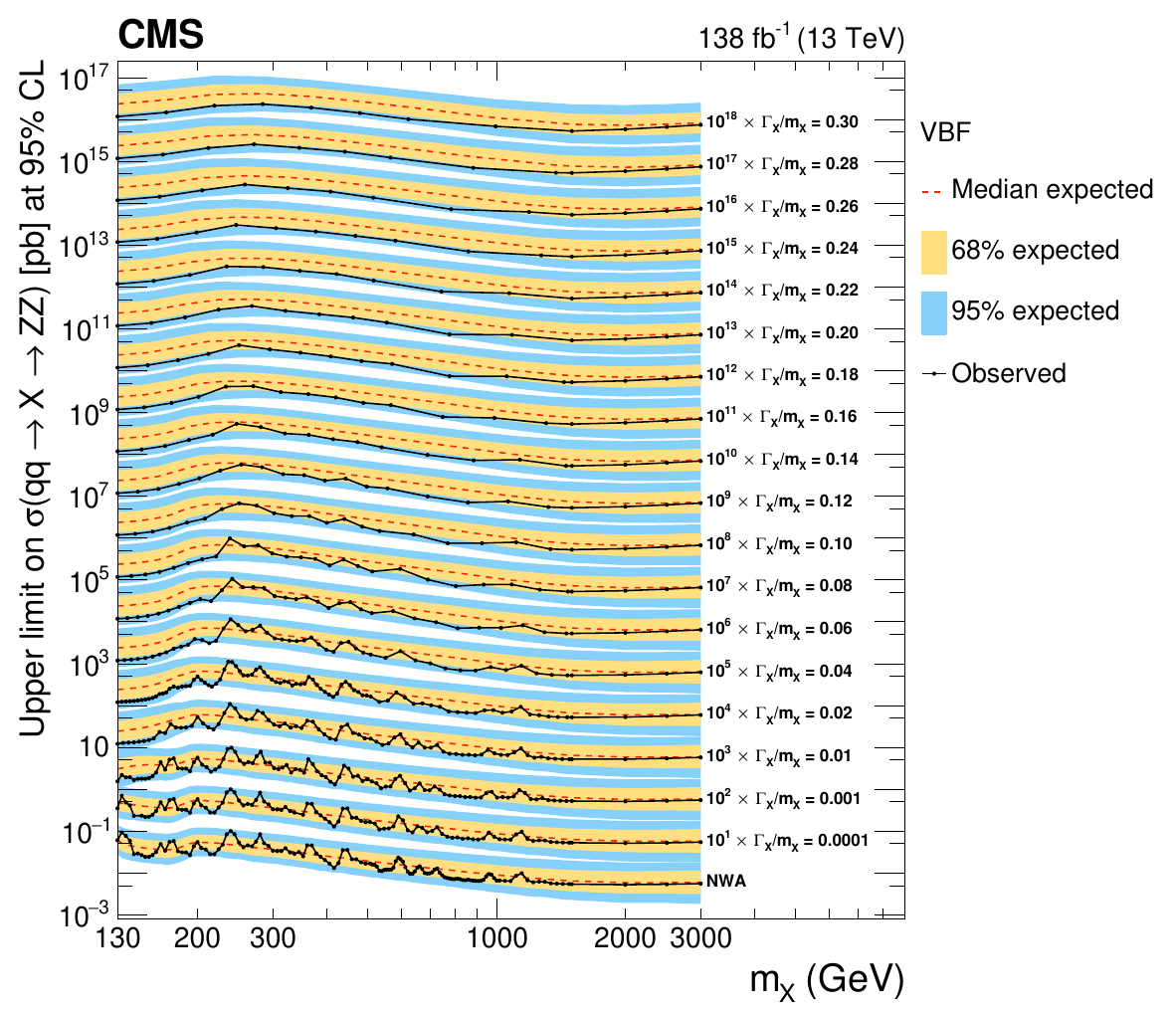}
    \caption{Observed and expected 95\% \CL upper limits on $\sigma(\Pp\Pp\to X\to ZZ)$ with $\MX$ from 130\GeV to 3\TeV and $\gammaX/\MX$ up to 30\%. The upper panel shows the results for pure ggF production, and the lower panel shows the results for pure VBF production.}
    \label{fig:scan2d}
\end{figure}

Results for variable $\MX$ and $\gammaX$ are shown in Fig.~\ref{fig:scan2d}. Each curve corresponds to an assumption on the width, from NWA to $\gammaX/\MX=0.3$. At small widths, the observed results fluctuate around the median expected limits; with the width growing, the observed results are smeared out, and excesses in the observed limits gradually disappear or move below the median expected results.

\clearpage

\section{Summary}
\label{sec:summary}

A search for a spin-0 resonance decaying to a pair of \PZ bosons in the four-lepton final state, where the leptons are muons or electrons, is performed at the CMS experiment. The data set used was collected in 2016--2018 and corresponds to an integrated luminosity of 138\fbinv. The searched-for resonance can be produced via gluon fusion or vector boson fusion. The mass of the sought resonance is scanned over a range from 130\GeV to 3\TeV, and different decay width assumptions are tested.

No significant excess over the standard model background expectation is observed. The largest fluctuation is seen at a mass of 137.8\GeV under the narrow-width assumption, reaching a global significance of 1.8 standard deviations. Upper limits at 95\% confidence level on the production cross section multiplied by the decay branching fraction of $\PX\to\PZ\PZ$ are set for various masses, decay widths, and production mechanisms. The exclusion limits range from 0.05--0.1\unit{pb} in the low-mass region to 0.005\unit{pb} in the high-mass region.

\begin{acknowledgments}
We congratulate our colleagues in the CERN accelerator departments for the excellent performance of the LHC and thank the technical and administrative staffs at CERN and at other CMS institutes for their contributions to the success of the CMS effort. In addition, we gratefully acknowledge the computing centers and personnel of the Worldwide LHC Computing Grid and other centers for delivering so effectively the computing infrastructure essential to our analyses. Finally, we acknowledge the enduring support for the construction and operation of the LHC, the CMS detector, and the supporting computing infrastructure provided by the following funding agencies: SC (Armenia), BMBWF and FWF (Austria); FNRS and FWO (Belgium); CNPq, CAPES, FAPERJ, FAPERGS, and FAPESP (Brazil); MES and BNSF (Bulgaria); CERN; CAS, MoST, and NSFC (China); MINCIENCIAS (Colombia); MSES and CSF (Croatia); RIF (Cyprus); SENESCYT (Ecuador); ERC PRG and PSG, TARISTU24-TK10 and MoER TK202 (Estonia); Academy of Finland, MEC, and HIP (Finland); CEA and CNRS/IN2P3 (France); SRNSF (Georgia); BMFTR, DFG, and HGF (Germany); GSRI (Greece); MATE and NKFIH (Hungary); DAE and DST (India); IPM (Iran); SFI (Ireland); INFN (Italy); MSIT and NRF (Republic of Korea); MES (Latvia); LMTLT (Lithuania); MOE and UM (Malaysia); BUAP, CINVESTAV, CONACYT, LNS, SEP, and UASLP-FAI (Mexico); MOS (Montenegro); MBIE (New Zealand); PAEC (Pakistan); MSHE, NSC, and NAWA (Poland); FCT (Portugal); MESTD (Serbia); MICIU/AEI and PCTI (Spain); MOSTR (Sri Lanka); Swiss Funding Agencies (Switzerland); MST (Taipei); MHESI (Thailand); TUBITAK and TENMAK (T\"{u}rkiye); NASU (Ukraine); STFC (United Kingdom); DOE and NSF (USA).

\begin{sloppy}
\setlength\emergencystretch{\hsize}
\hyphenation{Rachada-pisek} Individuals have received support from the Marie-Curie program and the European Research Council and Horizon 2020 Grant, contract Nos.\ 675440, 724704, 752730, 758316, 765710, 824093, 101115353, 101002207, 101001205, and COST Action CA16108 (European Union); the Leventis Foundation; the Alfred P.\ Sloan Foundation; the Alexander von Humboldt Foundation; the Science Committee, project no. 22rl-037 (Armenia); the Fonds pour la Formation \`a la Recherche dans l'Industrie et dans l'Agriculture (FRIA) and Fonds voor Wetenschappelijk Onderzoek contract No. 1228724N (Belgium); the Beijing Municipal Science \& Technology Commission, No. Z191100007219010, the Fundamental Research Funds for the Central Universities, the Ministry of Science and Technology of China under Grant No. 2023YFA1605804, the Natural Science Foundation of China under Grant No. 12535004, and USTC Research Funds of the Double First-Class Initiative No.\ YD2030002017 (China); the Ministry of Education, Youth and Sports (MEYS) of the Czech Republic; the Shota Rustaveli National Science Foundation (Georgia); the Deutsche Forschungsgemeinschaft (DFG), among others, under Germany's Excellence Strategy -- EXC 2121 ``Quantum Universe" -- 390833306, and under project number 400140256 - GRK2497; the Hellenic Foundation for Research and Innovation (HFRI), Project Number 2288 (Greece); the Hungarian Academy of Sciences, the New National Excellence Program - \'UNKP, the NKFIH research grants K 131991, K 138136, K 143460, K 143477, K 147557, K 146913, K 146914, K 147048, TKP2021-NKTA-64, and 2025-1.1.5-NEMZ\_KI-2025-00004, and MATE KKP and KKPCs Research Excellence and Flagship Research Groups grants (Hungary); the Council of Science and Industrial Research, India; ICSC -- National Research Center for High Performance Computing, Big Data and Quantum Computing, FAIR -- Future Artificial Intelligence Research, and CUP I53D23001070006 (Mission 4 Component 1), funded by the NextGenerationEU program, the Italian Ministry of University and Research (MUR) under Bando PRIN 2022 -- CUP I53C24002390006, PRIN PRIMULA 2022RBYK7T (Italy); the Latvian Council of Science; the Ministry of Science and Higher Education, project no. 2022/WK/14, and the National Science Center, contracts Opus 2021/41/B/ST2/01369, 2021/43/B/ST2/01552, 2023/49/B/ST2/03273, and the NAWA contract BPN/PPO/2021/1/00011 (Poland); the Funda\c{c}\~ao para a Ci\^encia e a Tecnologia (Portugal); the National Priorities Research Program by Qatar National Research Fund; MICIU/AEI/10.13039/501100011033, ERDF/EU, ``European Union NextGenerationEU/PRTR", projects PID2022-142604OB-C21, PID2022-139519OB-C21, PID2023-147706NB-I00, PID2023-148896NB-I00, PID2023-146983NB-I00, PID2023-147115NB-I00, PID2023-148418NB-C41, PID2023-148418NB-C42, PID2023-148418NB-C43, PID2023-148418NB-C44, PID2024-158190NB-C22, RYC2021-033305-I, RYC2024-048719-I, CNS2023-144781, CNS2024-154769 and Plan de Ciencia, Tecnolog{\'i}a e Innovaci{\'o}n de Asturias, Spain; the Chulalongkorn Academic into Its 2nd Century Project Advancement Project, the National Science, Research and Innovation Fund program IND\_FF\_68\_369\_2300\_097, and the Program Management Unit for Human Resources \& Institutional Development, Research and Innovation, grant B39G680009 (Thailand); the Eric \& Wendy Schmidt Fund for Strategic Innovation through the CERN Next Generation Triggers project under grant agreement number SIF-2023-004; the Kavli Foundation; the Nvidia Corporation; the SuperMicro Corporation; the Welch Foundation, contract C-1845; and the Weston Havens Foundation (USA).
\end{sloppy}
\end{acknowledgments}

\bibliography{auto_generated}
\cleardoublepage \appendix\section{The CMS Collaboration \label{app:collab}}\begin{sloppypar}\hyphenpenalty=5000\widowpenalty=500\clubpenalty=5000\cmsinstitute{Yerevan Physics Institute, Yerevan, Armenia}
{\tolerance=6000
A.~Hayrapetyan, V.~Makarenko\cmsorcid{0000-0002-8406-8605}, A.~Tumasyan\cmsAuthorMark{1}\cmsorcid{0009-0000-0684-6742}
\par}
\cmsinstitute{Institut f\"{u}r Hochenergiephysik, Vienna, Austria}
{\tolerance=6000
W.~Adam\cmsorcid{0000-0001-9099-4341}, L.~Benato\cmsorcid{0000-0001-5135-7489}, T.~Bergauer\cmsorcid{0000-0002-5786-0293}, M.~Dragicevic\cmsorcid{0000-0003-1967-6783}, C.~Giordano\cmsorcid{0000-0001-6317-2481}, P.S.~Hussain\cmsorcid{0000-0002-4825-5278}, M.~Jeitler\cmsAuthorMark{2}\cmsorcid{0000-0002-5141-9560}, N.~Krammer\cmsorcid{0000-0002-0548-0985}, A.~Li\cmsorcid{0000-0002-4547-116X}, D.~Liko\cmsorcid{0000-0002-3380-473X}, M.~Matthewman, J.~Schieck\cmsAuthorMark{2}\cmsorcid{0000-0002-1058-8093}, R.~Sch\"{o}fbeck\cmsAuthorMark{2}\cmsorcid{0000-0002-2332-8784}, M.~Shooshtari\cmsorcid{0009-0004-8882-4887}, M.~Sonawane\cmsorcid{0000-0003-0510-7010}, W.~Waltenberger\cmsorcid{0000-0002-6215-7228}, C.-E.~Wulz\cmsAuthorMark{2}\cmsorcid{0000-0001-9226-5812}
% ...
\par}
\vskip\cmsinstskip
\dag:~Deceased\\
$^{1}$Also at Yerevan State University, Yerevan, Armenia\\
$^{2}$Also at TU Wien, Vienna, Austria\\
% ...
\end{sloppypar}
\end{document}